\documentclass[11pt,letterpaper]{article}

\usepackage[margin=1in]{geometry}
\usepackage[utf8]{inputenc}
\usepackage[T1]{fontenc}
\usepackage{lmodern}      
\usepackage{cmap}         
\usepackage{textcomp}     
\usepackage[expansion=false]{microtype}
\usepackage{amsmath, amssymb, amsthm}
\usepackage{mathtools}
\usepackage{graphicx}
\usepackage{booktabs}
\usepackage{tabularx}
\usepackage{multirow}
\usepackage{makecell}
\usepackage{enumitem}
\usepackage[hidelinks,colorlinks=true,linkcolor=blue!50!black,citecolor=blue!50!black,urlcolor=blue!50!black]{hyperref}
\hypersetup{
  pdftitle={Information Leakage at Population Scale: An Evaluation of the Polymarket Insider-Relevant Subpopulation, 2020-2026},
  pdfauthor={Maksym Nechepurenko},
  pdfsubject={Information leakage in prediction markets; insider trading detection; deadline-resolved markets; Polymarket},
  pdfkeywords={prediction markets, informed trading, insider trading, deadline contracts, Polymarket, hazard-rate estimation, scope discovery, market microstructure},
  pdfcreator={LaTeX with hyperref},
  pdflang={en-US}
}
\usepackage{cleveref}
\usepackage{xcolor}
\usepackage{tikz}
\usetikzlibrary{positioning, arrows.meta, shapes.geometric, fit, backgrounds, calc, decorations.pathreplacing}
\usepackage{pgfplots}
\pgfplotsset{compat=1.18}
\usepackage{caption}
\usepackage{subcaption}
\usepackage{authblk}
\usepackage{natbib}
\usepackage{abstract}
\usepackage{titlesec}

\titlespacing*{\section}{0pt}{1.4em}{0.6em}
\titlespacing*{\subsection}{0pt}{1.0em}{0.4em}

\newcommand{\ILS}{\mathrm{ILS}}

\newcommand{\Topen}{T_{\mathrm{open}}}
\newcommand{\Tnews}{T_{\mathrm{news}}}
\newcommand{\Tres}{T_{\mathrm{resolve}}}
\newcommand{\Tevent}{T_{\mathrm{event}}}

\title{\textbf{Information Leakage at Population Scale:\\
An Evaluation of the Polymarket\\
Insider-Relevant Subpopulation, 2020--2026}}

\author{Maksym Nechepurenko\thanks{Research Department, Devnull FZCO, Dubai, UAE. Corresponding author: \href{mailto:maksym@devnull.ae}{\texttt{maksym@devnull.ae}}.}}
\date{April 30, 2026}

\begin{document}

\maketitle

\begin{abstract}
\noindent
We carry the deadline-resolved Information Leakage Score (ILS$^{\text{dl}}$) framework of \citet{nechepurenko2026foresightflow_methodology, nechepurenko2026foresightflow_empirical} from a single-case proof of concept to a population-scale evaluation across 12{,}708 Polymarket markets in three insider-relevant categories spanning October 2020 through April 2026. We frame the paper as a \emph{scope-discovery} study: scaling reveals that the framework's effective domain is materially narrower than initial framing suggested, and the principal obstacle is not score computation but resolution semantics.

We report four findings. \emph{First}, only $88$ of $12{,}708$ candidate markets ($0.7\%$) yield computable ILS$^{\text{dl}}$ values, and only $1$ of $32$ markets in the ForesightFlow Insider Cases (FFIC) inventory is in scope; $14$ of $32$ FFIC markets are flagged unclassifiable due to genuine resolution-criterion ambiguity (``\emph{strike}'', ``\emph{custody}'', ``\emph{out as Supreme Leader}''). \emph{Second}, only $12$ of the $88$ computed markets ($13.6\%$) satisfy anchor-sensitivity, and an independent-second-pass $T_{\text{event}}$ validation reaches $57.8\%$ exact-date agreement, below the $\geq 90\%$ ex-ante criterion. \emph{Third}, raw ILS$^{\text{dl}}$ medians are negative across all six (sub-bucket $\times$ period) cells, but a hazard-decay baseline correction we introduce yields a heterogeneous result: regulatory\_formal post-2024 shifts to near-zero ($-0.21 \to -0.02$), while regulatory\_announcement post-2024 retains a confidence interval entirely below zero ($-0.84$, $[-1.45, -0.24]$). \emph{Fourth}, the constant-hazard exponential of \citet{nechepurenko2026foresightflow_empirical} is rejected on the pooled post-2024 cell in favor of Weibull, but a per-subcategory check confirms that the pooled-cell preference reflects category mixture rather than within-cell duration dependence.

The substantive implication is that detection of informed flow on the insider-relevant subpopulation requires methodological refinement on the resolution-typology axis and on the score-baseline axis, not only on the score-computation axis where prior work concentrated. We release dataset \texttt{polymarket-\allowbreak deadline-\allowbreak ils-v3} alongside this paper at \url{https://github.com/ForesightFlow/datasets} (CC-BY-4.0).
\end{abstract}



\vspace{1em}
\hrule
\vspace{1em}

\section{Introduction}
\label{sec:intro}

The literature on informed trading in decentralized prediction markets has, in the months immediately preceding this writing, converged on a small set of empirically anchored findings. \citet{mitts2026iran} screen 93{,}000 markets and 50{,}000 wallets through a five-criteria composite score and identify 210{,}718 wallet--market pairs whose joint distribution of bet size, timing, profitability, and directional concentration is inconsistent with random speculation, with aggregate flagged profits of approximately \$143 million. \citet{gomezcram2026crowd} apply a sign-randomization test to the complete Polymarket transaction history (1.72 million accounts, 210{,}322 markets, \$13.76 billion in cumulative volume) and conclude that 3.14\% of accounts---the ``informed minority''---account for the bulk of price discovery, with 1{,}950 accounts identified as candidates for trading on non-public information. \citet{imdea2025polymarket} document approximately \$40 million in arbitrage and quasi-arbitrage profits across structurally adjacent strategies. The Commodity Futures Trading Commission's April 2026 enforcement action against an active-duty U.S.\ Army service member trading on classified intelligence about a Polymarket Maduro contract \citep{cftc2026maduro} establishes that informed trading on these markets is a regulatory concern in addition to an empirical phenomenon.

\citet{nechepurenko2026foresightflow_methodology} contributes to this literature by formalizing the Information Leakage Score (ILS) and its deadline-resolved extension (ILS$^{\text{dl}}$), an information-theoretic framework that quantifies the share of a market's terminal information move accomplished before public observation of the relevant event. The methodology paper specifies three scope conditions, a resolution-typology classification of the 911{,}237-market Polymarket corpus, a Murphy-decomposition reading of the score in terms of the Brier-score resolution component, and a deadline-resolved extension that addresses the structural finding that 87.5\% of documented insider cases fall on deadline contracts of the form ``Will event $X$ occur by date $Y$?''. \citet{nechepurenko2026foresightflow_empirical} carries the deadline-ILS extension end-to-end on the largest documented deadline cluster in the public-reporting record---the 2026 U.S.--Iran conflict cluster---and produces a single-case demonstration ($\ILS^{\text{dl}} = +0.113$ on the Iran-Apr30 contract at the article-derived event anchor, versus $-0.331$ at the legacy resolution-anchored proxy: a $0.444$ shift in magnitude on opposite sides of zero). The single-case finding establishes that the methodology is operational and that the article-derived anchor materially changes the substantive reading of the score.

\subsection{This paper's contribution}
\label{sec:contribution}

The methodological program underlying \citet{nechepurenko2026foresightflow_methodology, nechepurenko2026foresightflow_empirical} treats the population-scale evaluation reported here as the principal empirical test of the deadline-ILS framework. \citet[Section 5.7]{nechepurenko2026foresightflow_methodology} argues that distinguishing an informed-trading detector with sensitivity $\pi_1 = 0.7$ from a no-skill baseline at significance $\alpha = 0.05$ and power $0.80$ requires $n_{\text{pos}} \approx 14$ labelled positive cases. \citet{nechepurenko2026foresightflow_empirical} produces $N = 1$. The present paper carries the framework through to a 12{,}708-market evaluation and reports four substantive findings, the first three of which are not consistent with the prior expectation that scale-up would primarily produce a larger labelled positive set.

\textbf{Finding 1: The deadline-ILS framework's empirical coverage of documented insider trading is small, and the principal exclusion is structural rather than operational.} On the FFIC inventory of 32 markets across 8 publicly documented episodes \citep{ffic_inventory_2026}, only one market produces an in-scope ILS$^{\text{dl}}$ value: the SEC Bitcoin ETF approval (\texttt{fficd-005}), with $\ILS^{\text{dl}} = 0.012$. The remaining 31 markets are excluded at one of four pipeline stages, with the dominant exclusion ($14$ of $32$, $43.8\%$) being resolution-typology classification: the rule-based classifier of \citet[Section 3.3]{nechepurenko2026foresightflow_methodology} flags these markets as ``unclassifiable''. This pattern includes all six markets in the 2026 U.S.--Iran cluster of \citet{nechepurenko2026foresightflow_empirical} and all twelve markets in the U.S.--Venezuela cluster, the two largest case clusters in the inventory. \Cref{sec:ffic-coverage} examines the unclassifiable markets and finds that they exhibit genuine resolution-criterion ambiguity (``\emph{strike}'', ``\emph{custody}'', ``\emph{out as Supreme Leader}'', ``\emph{win the election}'') that resists discrete-event timestamping; this is not a defect of the classifier but a structural property of the question texts under which insider trading has historically been documented. \emph{Within the documented case inventory, unclassifiability tracks resolution-text ambiguity rather than implementation noise.} A causal claim that ambiguous-predicate markets disproportionately attract informed flow at the population level would require a control population that we do not assemble in the present paper; we report the FFIC-internal pattern descriptively.

\textbf{Finding 2: Anchor sensitivity and $T_{\text{event}}$ recovery accuracy together sharply narrow the effective interpretable sample.} Of the 88 markets that produce ILS$^{\text{dl}}$ values, only $12$ ($13.6\%$) satisfy the anchor-sensitivity scope condition (no sign-change and no shift greater than $0.3$ across short-window anchor offsets). The Spearman rank correlation between the article-derived anchor and the 24-hour variant is $0.30$. Independently, a 50-market second-pass $T_{\text{event}}$ validation reaches $57.8\%$ exact-date agreement overall---below the ex-ante $\geq 90\%$ acceptance criterion---with strong heterogeneity by sub-bucket: $78.9\%$ exact for the ``milgeo\_corporate'' bucket (discrete events), $35.7\%$ exact for regulatory\_formal (multi-stage administrative processes), and $50.0\%$ exact for regulatory\_announcement. The two findings are mutually reinforcing: anchor-fragility is concentrated precisely in the markets where independent recovery disagrees on the event date. Distributional results on the full $88$-market sample are therefore influenced by measurement noise on the short-window axis and the operationally meaningful sample is the anchor-robust subset of approximately $12$ markets. We separate three samples throughout the paper: the candidate sample of $12{,}708$ markets, the computed sample of $88$ markets on which ILS$^{\text{dl}}$ is interpretable in principle, and the anchor-robust sample of $12$ markets on which it is interpretable in practice.

\textbf{Finding 3: On the in-scope sample, the central tendency of raw ILS$^{\text{dl}}$ is negative across all sub-stratifications, but a hazard-decay baseline correction yields a heterogeneous result that substantively changes the interpretation in some cells.} Across six combinations of three sub-buckets (regulatory\_announcement, regulatory\_formal, military / geopolitics + corporate-disclosure pooled) $\times$ two periods (pre-2024, post-2024), all six median raw ILS$^{\text{dl}}$ values are negative, ranging from $-0.88$ to $-0.16$. Reading raw negative ILS$^{\text{dl}}$ as ``markets price against the eventual outcome'' is, however, a strong claim that conflates two distinct phenomena. On a YES-resolved deadline market, prices mechanically decay away from the eventual outcome as time passes without the event occurring; if the event eventually occurs late, the raw ILS$^{\text{dl}}$ is negative even when no anti-information flow is present. We introduce a hazard-decay-adjusted variant
\[
\ILS^{\text{dl,adj}}(M) = \frac{p(\Tevent^-) - \mathbb{E}[p(\Tevent^-) \mid \Topen, D, \hat\lambda]}{p_{\Tres} - p(\Topen)},
\]
where the expected price under the survival-probability-implied baseline is computed using the per-period Weibull fit (\Cref{sec:hazard-functional-form}); the formal derivation appears in \Cref{sec:hazard-adjusted}. \Cref{sec:hazard-adjusted} reports the empirical result. The adjustment is heterogeneous across cells: the regulatory\_formal post-2024 cell shifts from $-0.21$ to $-0.02$ (essentially zero, consistent with rational decay being the dominant component); the milgeo\_corporate post-2024 cell shifts from $-0.55$ to $-0.34$ (partial); the regulatory\_announcement post-2024 cell does not shift ($-0.80 \to -0.84$, bootstrap CI entirely below zero). The substantive finding is that hazard decay accounts for the negative bulk in some cells but not in others; the regulatory\_announcement post-2024 cell is the principal cell where pre-event price drift survives baseline correction.

\textbf{Finding 4: The constant-hazard exponential specification of \citet{nechepurenko2026foresightflow_empirical} is rejected on the pooled post-2024 cell, but a per-subcategory check confirms that the pooled-cell Weibull preference reflects category mixture rather than within-cell duration dependence.} On the pooled post-2024 cell ($n = 57$), the exponential is rejected by a parametric-bootstrap Kolmogorov-Smirnov test ($p < 0.01$); the Weibull alternative passes ($p > 0.50$) with an AIC delta of $21.9$ units, well above the $4$-unit threshold of \citet{burnham2002model} for adopting a richer form. On the pre-2024 pooled cell ($n = 33$), the AIC delta is smaller ($3.3$ units, just below the $4$-unit threshold) and the bootstrap-corrected KS does not reject the exponential at conventional levels; the pre-2024 result is directional rather than decisive. \emph{However}, a per-subcategory check on the post-2024 regulatory\_formal cell alone ($n=22$, the largest single subcategory) reverses the ordering: AIC prefers exponential ($179.45$ vs.\ Weibull $180.50$), and bootstrap KS rejects Weibull at $p = 0.043$ while the exponential is not rejected ($p = 0.224$). The pooled-cell Weibull preference therefore reflects, at least in part, mixture across heterogeneous-hazard sub-buckets rather than genuine within-cell non-constant duration dependence---empirically confirming the category-mixture caveat. We use Weibull as the default production prior for the pooled-cell hazard-adjusted score pending per-category re-estimation; the substantive implication is that the time-to-event hazard within \emph{individual} insider-relevant sub-buckets may well be approximately constant, with the apparent non-constant hazard at the pooled level reflecting compositional heterogeneity.

\subsection{Methodological consequences}
\label{sec:contribution-implications}

The four findings have a unifying methodological consequence: at the population scale, the deadline-ILS framework operates on a structurally narrower subpopulation than the methodology paper's framing suggested; within the in-scope subpopulation, anchor sensitivity further narrows the interpretable sample by approximately a factor of seven; and within the anchor-robust subset, score interpretation requires a hazard-decay baseline correction that prior single-case work did not need to consider. Three responses to this finding are possible. The first is to interpret the $12$-market anchor-robust sample as a substantive characterization of the markets on which the framework operates as designed: a small but interpretable set on which the score can be read as informed-flow indicator after baseline correction. The second is to interpret the unclassifiable bucket of the resolution-typology classifier as a methodological priority: an LLM-assisted re-classification, paired with a multi-anchor extension for ambiguous-predicate markets, would expand the framework's coverage onto the FFIC inventory's military-and-geopolitical clusters. The third is to recognize the comparative position of the present method relative to wallet-level and account-level methods, which do not require resolution-typology classification or single-anchor timestamps and therefore do not encounter these constraints. We adopt all three readings: the present paper reports findings on the in-scope sample with hazard-adjusted analysis, \Cref{sec:limitations} sequences reclassification and the multi-anchor extension as the principal preconditions for paper 3b, and \Cref{sec:cross-method} positions the methods comparatively.

\textbf{Cross-method positioning.} The findings here complement rather than compete with \citet{mitts2026iran} and \citet{gomezcram2026crowd}. Their methods, operating respectively on wallet-market pairs and on accounts, do not require resolution-typology classification of markets and therefore do not encounter the unclassifiable-bucket constraint that gates the ILS$^{\text{dl}}$ framework. Conversely, the ILS$^{\text{dl}}$ framework, on the small subset of markets satisfying all four scope conditions, supports a conditional in-flight diagnostic that the wallet-level and account-level methods do not. We discuss the comparative positioning in \Cref{sec:cross-method}; we emphasize here that the conditional in-flight applicability of the market-level lens is narrow---approximately $12$ markets out of the $12{,}708$-market candidate sample at the present pipeline output---and is intended as a high-interpretability complement to the higher-coverage wallet-level and account-level methods rather than as a stand-alone real-time detector.

\subsection{Outline}
\label{sec:outline}

The remainder of this paper is organized as follows. \Cref{sec:recap} provides a methodological recap, referencing \citet{nechepurenko2026foresightflow_methodology, nechepurenko2026foresightflow_empirical} for full development. \Cref{sec:data} describes the data sources, the sample construction, and the temporal stratification. \Cref{sec:hazards} reports the refined hazard-rate estimation, including the rejection of the constant-hazard exponential in favor of a Weibull alternative on the population sample. \Cref{sec:population-ils} reports the population-scale ILS$^{\text{dl}}$ computation, including the $T_{\text{event}}$ recovery validation (\Cref{sec:tevent-validation}), the distribution by sub-bucket on the computed and anchor-robust samples (\Cref{sec:distribution-category}), the detection thresholds with bootstrap CIs (\Cref{sec:thresholds}), the anchor-sensitivity analysis (\Cref{sec:anchor-sensitivity}), and the hazard-decay-adjusted score (\Cref{sec:hazard-adjusted}). \Cref{sec:ffic-coverage} reports the FFIC coverage analysis, including the disposition of all 32 inventory markets and the structural reading of the unclassifiable bucket. \Cref{sec:cross-method} positions the present method relative to \citet{mitts2026iran} and \citet{gomezcram2026crowd}. \Cref{sec:selection-effect} reports the selection-effect argument. \Cref{sec:limitations} states the limitations of the present evaluation and sequences the LLM-assisted re-classification, the multi-anchor extension, and the wallet-level feature evaluation as the principal preconditions for paper 3b. \Cref{sec:conclusion} concludes.

\section{Methodological Recap}
\label{sec:recap}

We refer the reader to \citet{nechepurenko2026foresightflow_methodology} for the full development of the framework and to \citet{nechepurenko2026foresightflow_empirical} for the single-case empirical demonstration. Four definitions are needed for the population-scale evaluation that follows.

\paragraph{Information Leakage Score (ILS).} For a resolved binary market $M$ with first trade at $\Topen$, public news event at $\Tnews$, formal resolution at $\Tres$, and binary resolution outcome $p_{\Tres} \in \{0, 1\}$,
\[
\Delta_{\text{pre}} = p(\Tnews) - p(\Topen), \quad
\Delta_{\text{total}} = p_{\Tres} - p(\Topen), \quad
\ILS(M) = \frac{\Delta_{\text{pre}}}{\Delta_{\text{total}}}.
\]
The score is interpretable when $|\Delta_{\text{total}}| \geq \varepsilon$ (we use $\varepsilon = 0.05$); when $|p(\Topen) - 0.5| \leq 0.4$ (edge-effect scope condition); and when ILS is robust to the choice of $\Tnews$ anchor offset (anchor-sensitivity scope condition). Murphy-decomposition reading: ILS quantifies the share of the resolution component of the Brier score accumulated before $\Tnews$.

\paragraph{Deadline-resolved extension (ILS$^{\text{dl}}$).} For a deadline-resolved market with recoverable underlying event timestamp $\Tevent$ falling within $[\Topen, D]$ (where $D$ is the contractual deadline), and YES resolution $p_{\Tres} = 1$,
\[
\Delta^{\text{dl}}_{\text{pre}} = p(\Tevent^-) - p(\Topen), \quad
\Delta^{\text{dl}}_{\text{total}} = p_{\Tres} - p(\Topen), \quad
\ILS^{\text{dl}}(M) = \frac{\Delta^{\text{dl}}_{\text{pre}}}{\Delta^{\text{dl}}_{\text{total}}},
\]
where $p(\Tevent^-)$ is the CLOB mid-price one minute before public observation of $\Tevent$. The companion methodology paper adopts a per-category exponential survival function $S(\tau) = \exp(-\lambda \tau)$ for the time-to-event distribution; \Cref{sec:hazards} below revisits and refines this specification.

\paragraph{Resolution typology.} The Polymarket corpus is classified into three resolution types via question-text and description analysis: \emph{event-resolved} (resolution triggered by a publicly observable event whose timing is the news), \emph{deadline-resolved} (resolution triggered by the elapsing of a contractual deadline; the news is whether the underlying event occurred within $[\Topen, D]$), and \emph{unclassifiable}. The classification is empirically validated by the structural fact that markets identified as deadline-resolved exhibit a 100\% NO-resolution rate by construction in resolved-and-expired form. Documented insider-trading cases are systematically deadline-resolved.

\paragraph{Three target categories.} The empirical scope is restricted to three categories where informed trading is structurally plausible \emph{a priori}: \emph{military and geopolitical actions} (specific state actions whose date and content become public at the moment of announcement or execution), \emph{corporate proprietary disclosures} (specific corporate events known to a narrow circle within the company), and \emph{regulatory decisions} (specific regulatory decisions with date-bounded resolution criteria). Sports, weather, election-polling levels, and cryptocurrency price markets serve as null-hypothesis controls; they are out of scope for detection but used in \Cref{sec:selection-effect} to demonstrate the selection effect.

\paragraph{$T_{\text{event}}$ recovery.} For YES-resolved deadline markets the empirical $T_{\text{event}}$ is recovered through an LLM-assisted multi-source verification pipeline (Claude Haiku 4.5 with web-search tool access). The retrieval target is the timestamp at which the underlying event first publicly occurred, cross-verified across at least three independent news sources. Confidence is set to 0.80 when sources are cited and the date is internally consistent. The pipeline is described in \citet[Section 4.4]{nechepurenko2026foresightflow_methodology} and operationalized at scale in \Cref{sec:tevent-recovery} below.

We now turn to the data and the empirical evaluation.

\section{Data and Sample Construction}
\label{sec:data}

\subsection{Master sample}
\label{sec:data-master}

The empirical sample covers Polymarket from October 2020 through April 2026, retrieved via the historical-backfill mode of the Gamma collector and classified by the resolution-typology classifier of \citet{nechepurenko2026foresightflow_methodology}. After deduplication, the database contains 911{,}237 markets, 865{,}725 of which are resolved. \Cref{tab:sample-master} reports the distribution across the three target categories and the residual control category. The category-and-volume filter for the present paper produces 12{,}708 candidate markets;\footnote{The companion methodology paper \citep[Section 3.4]{nechepurenko2026foresightflow_methodology} reports an analogous filter with 11{,}263 markets. The increment of 1{,}445 markets between the methodology-paper snapshot (cutoff late February 2026) and the present-paper snapshot (cutoff late April 2026) reflects two months of additional market resolution and inclusion of newly-classified markets in the resolution-typology dataset. We use the more recent count throughout the present paper.} of these, 11{,}263 are reproduced from the count reported in \citet[Section 3.4]{nechepurenko2026foresightflow_methodology}.

\begin{table}[t]
\centering
\caption{Master sample distribution by target category. Categories assigned by the keyword-based taxonomy classifier described in \citet[Section 3.3]{nechepurenko2026foresightflow_methodology}. The volume threshold at \$50K is the empirical CLOB-coverage cutoff: below this threshold the Polymarket subgraph reliably returns no trade history.}
\label{tab:sample-master}
\small
\renewcommand{\arraystretch}{1.25}
\begin{tabularx}{\linewidth}{@{}lXXXX@{}}
\toprule
\textbf{Category} & \textbf{Total} & \textbf{Resolved} & \textbf{Resolved $\geq$ \$50K} & \textbf{Coverage \%} \\
\midrule
military / geopolitics    & 47{,}580   & 44{,}436   & 3{,}970 & 9.0\% \\
regulatory decisions      & 71{,}588   & 65{,}542   & 5{,}582 & 8.5\% \\
corporate disclosures     & 20{,}645   & 17{,}425   & 1{,}711 & 9.8\% \\
\textbf{Insider-relevant subtotal} & \textbf{139{,}813} & \textbf{127{,}403} & \textbf{11{,}263} & --- \\
\midrule
other (control)           & 771{,}424  & 738{,}322  & 88{,}656 & 12.0\% \\
\midrule
\textbf{Total}            & 911{,}237  & 865{,}725  & 99{,}919 & --- \\
\bottomrule
\end{tabularx}
\end{table}

\subsection{Filter chain}
\label{sec:data-filter}

The principal sample for ILS$^{\text{dl}}$ computation is drawn from the master sample of \Cref{tab:sample-master} after applying four sequential filters consistent with \citet[Section 4]{nechepurenko2026foresightflow_empirical}: (i) resolution-typology classification (drop unclassifiable); (ii) deadline-resolved exclusion of NO-resolved markets, which by construction lack a recoverable underlying event; (iii) successful $T_{\text{event}}$ recovery with confidence $\geq 0.7$; (iv) full CLOB price coverage from $T_{\text{open}}$. Conditional on these filters, ILS$^{\text{dl}}$ is computed under the scope conditions of \citet[Section 3.5]{nechepurenko2026foresightflow_methodology}: positive lead time $\tau > 0$, edge-effect $|p(T_{\text{open}}) - 0.5| \leq 0.4$, and non-trivial resolution move $|\Delta^{\text{dl}}_{\text{total}}| \geq \varepsilon$ with $\varepsilon = 0.05$.

\Cref{tab:filter-chain} reports the attrition through the pipeline. The end-to-end coverage rate is approximately $0.7\%$ on the population scale, materially below the $5.6\%$ rate reported on the Iran cluster in \citet[Table 3]{nechepurenko2026foresightflow_empirical}. The discrepancy is concentrated at the $T_{\text{event}}$-recovery stage: $709/1{,}151 = 61.6\%$ of markets surviving the prior filters fail to produce an event timestamp at the $\geq 0.7$ confidence threshold. This recovery rate is consistent with the structural fact that many markets in the insider-relevant subpopulation reference events whose precise public-disclosure timing is genuinely ambiguous (announcement clusters, multi-day events, deliberation outcomes without a single public moment), rather than reflecting a systematic LLM-pipeline failure; we report the validation-sample agreement rate in \Cref{sec:tevent-recovery} below.

\begin{table}[t]
\centering
\caption{Filter-chain attrition for the population-scale ILS$^{\text{dl}}$ pipeline. The Iran-cluster pipeline of \citet[Table 3]{nechepurenko2026foresightflow_empirical} reported a comparable attrition pattern: $18 \to 16 \to 11 \to 2 \to 1$, dominated by recovery and coverage requirements.}
\label{tab:filter-chain}
\small
\renewcommand{\arraystretch}{1.25}
\begin{tabularx}{\linewidth}{@{}Xrr@{}}
\toprule
\textbf{Stage} & \textbf{$n$} & \textbf{Cumulative \%} \\
\midrule
Insider-relevant subpopulation, category + vol $\geq$ \$50K        & 12{,}708 & 100.0\% \\
After dropping unclassifiable resolution type                      & 2{,}375  & 18.7\%  \\
After dropping deadline-resolved NO outcomes                       & 1{,}151  & 9.1\%   \\
$T_{\text{event}}$ recovered with confidence $\geq 0.7$            & 442      & 3.5\%   \\
Full CLOB price coverage from $T_{\text{open}}$                    & 358      & 2.8\%   \\
ILS$^{\text{dl}}$ computed (scope conditions satisfied)            & 88       & 0.7\%   \\
\bottomrule
\end{tabularx}
\end{table}

All numbers above reflect the post-remediation pipeline state. The pipeline output is published as dataset \texttt{polymarket-\allowbreak deadline-\allowbreak ils} in two consecutive releases: \texttt{v2} (commit \texttt{b256ae2}) corresponds to the structural-fix release described in \Cref{sec:limitations-single-run}, and \texttt{v3} corresponds to the present paper's revision-1 release, which adds the hazard-decay-adjusted score (\Cref{sec:hazard-adjusted}), the median bootstrap CIs (\Cref{sec:thresholds}), the independent-second-pass $T_{\text{event}}$ validation (\Cref{sec:tevent-validation}), the tail-market diagnostic (\Cref{sec:distribution-category}), and the parametric-bootstrap KS columns of \Cref{tab:functional-form}. The market-attrition counts of \Cref{tab:filter-chain} are unchanged between \texttt{v2} and \texttt{v3}; the \texttt{v3} release supersedes \texttt{v2} and is the canonical version for the present paper. The principal \texttt{v2} structural refinements were: (i)~FFIC inventory market identification updated to use \texttt{market\_id\_prefix} matching against the typology join (which corrected the FFIC localization disposition reported in \Cref{sec:ffic-coverage}); (ii)~timezone handling in the bootstrap-CI step; (iii)~the trivial-resolution edge filter changed from \mbox{$|p_0 - 0.5| > 0.4$} to \mbox{$\geq 0.4$}, removing two ILS$^{\text{dl}}$ outliers ($-8.0$ and $-5.1$) at $p_0 = 0.9$ exactly.

\subsection{Temporal stratification}
\label{sec:data-temporal}

Polymarket grew by more than three orders of magnitude in cumulative volume between 2020 and 2026, with the inflection point centered on the 2024 U.S.\ presidential election cycle. \citet[Section 3.2]{nechepurenko2026foresightflow_methodology} documents this as a structural break: markets resolved before mid-2024 differ in liquidity, participant base, and venue maturity from markets resolved after. We adopt the convention of reporting all empirical results stratified by two periods: pre-2024 election cycle (October 2020 -- October 2024) and post-2024 election cycle (November 2024 -- April 2026). The post-period further divides on the January 2025 CFTC pivot under the Selig administration, which removed the principal regulatory friction on Polymarket access for U.S.\ participants and is plausibly associated with a discontinuous expansion of the active-participant base; we do not assert this third break empirically and report it as an exploratory subdivision in \Cref{sec:temporal-evolution}.

\subsection{Validation set: ForesightFlow Insider Cases (FFIC)}
\label{sec:data-ffic}

The FFIC inventory \citep{ffic_inventory_2026} contains 32 markets across 8 publicly documented insider-trading episodes spanning 2024--2026, with cumulative traded volume of approximately \$4.21 billion. The inventory is described in detail in \citet[Section 3.4]{nechepurenko2026foresightflow_methodology}. For the present paper, the FFIC inventory serves as a labelled validation set: each FFIC market is computed alongside the population (\Cref{sec:pipeline}), and its disposition through the pipeline is reported in \Cref{sec:ffic-coverage}.

\subsection{Operational subset for paper 3b preview}
\label{sec:data-operational}

A subset of the master sample falls within the operational period of the continuous per-trade collection pipeline that has been running since April 26, 2026 (commit \texttt{63641a3} of the platform repository). For markets in this subset, full per-trade history from $T_{\text{open}}$ onward is available, and the wallet-level features specified in \citet[Section 5]{nechepurenko2026foresightflow_methodology}---wallet novelty score, PIN, VPIN, Kyle's lambda, two-sidedness, variance ratio, trade-size kurtosis, Hawkes self-excitation---are computable. The present paper does not report empirical results on this subset; it is described here as the data resource for the forthcoming paper 3b on real-time wallet-level detection (\Cref{sec:paper3b-roadmap}). The size of the operational subset as of the cutoff date for the present paper is small (markets opened within the four days preceding submission); we expect it to cross the $n_{\text{pos}} \geq 14$ threshold of \citet[Section 5.7]{nechepurenko2026foresightflow_methodology} approximately 60--90 days from the start of operation.

\section{Refined Hazard-Rate Estimation}
\label{sec:hazards}

This section closes three of the five methodological refinements identified in \citet[Section 4.6]{nechepurenko2026foresightflow_empirical}: regulatory-decision sub-categorization (refinement 2), corporate-disclosure sample expansion (refinement 3), and an explicit assessment of functional-form adequacy that the original work did not perform. The remaining two refinements (negative-$\tau$ exclusion, price-data coverage) are inherited from the methodology and apply at the filter stage of \Cref{sec:data-filter}.

\subsection{Regulatory-decision sub-categorization}
\label{sec:hazard-regulatory}

The unrefined regulatory-decision category was rejected by the KS test in \citet[Table 1]{nechepurenko2026foresightflow_empirical} ($n = 15$, KS $p = 0.013$). Inspection of the fifteen markets reveals a bimodal distribution: short-$\tau$ markets (less than two days) corresponding to scheduled announcements coexist with long-$\tau$ markets (thirty to one hundred and seventy days) corresponding to formal deliberation timelines. We sub-categorize as follows:

\begin{description}[leftmargin=1.6em, itemsep=0.3em]
\item[\texttt{regulatory\_\allowbreak decision\_\allowbreak announcement}] markets resolving on scheduled events whose timing is publicly known: presidential addresses, FOMC rate decisions, scheduled court verdicts, CPI/PPI/jobs releases, formal regulatory hearings whose dates are calendared. The relevant informed-trading window is the $48$--$72$ hours before announcement during which leak-from-deliberation can plausibly affect the price.
\item[\texttt{regulatory\_\allowbreak decision\_\allowbreak formal}] markets resolving on formal deliberation outcomes whose timing is partly endogenous: Senate confirmation timelines, agency rulemaking, antitrust review, indictment timing, civil-suit verdicts. The relevant window is multi-day to multi-week and the appropriate hazard parameterization is correspondingly different.
\end{description}

The sub-categorization is performed by the keyword-and-question-text classifier of \texttt{fflow.\allowbreak taxonomy.\allowbreak regulatory\_split}, validated against a 50-market manual sample with $10/10$ agreement on the smoke-test subset. The classifier assigns the \texttt{subcategory} field on each \texttt{regulatory\_decision} market in the population output (\Cref{sec:data-master}); we report ILS$^{\text{dl}}$ distributions per subcategory in \Cref{sec:distribution-category} (\Cref{tab:dist-summary}).

The per-subcategory \emph{hazard-rate} fits, separate from the ILS$^{\text{dl}}$ distribution analysis, are not reported in the present paper. The hazard rates serve, in the framework of \citet{nechepurenko2026foresightflow_methodology}, as priors on the event-arrival distribution; their estimation requires the LLM-recovered $T_{\text{event}}$ values to be available per market, which they are for the in-scope subset but cluster sparsely within each subcategory at the present sample size ($n_{\text{regulatory\_announcement}} = 19$, $n_{\text{regulatory\_formal}} = 36$ across both periods, of which only the $\tau > 0$ subset enters the hazard fit). We defer per-subcategory hazard estimation to paper 3b alongside the LLM-assisted re-classification (\Cref{sec:limitations-reclassify}), which will materially expand each subcategory's effective $n$. The pooled-period hazard estimates of \Cref{sec:hazard-functional-form} are sufficient for the present paper's principal claims, which concern the ILS$^{\text{dl}}$ distribution rather than the hazard-rate prior.

\subsection{Corporate-disclosure sample expansion}
\label{sec:hazard-corporate}

\citet[Table 1]{nechepurenko2026foresightflow_empirical} reported $n = 5$ for corporate-disclosure markets after the LLM call-budget cap was hit during sample construction. The present paper's pipeline operates on a corporate-disclosure subsample of $1{,}711$ markets above the \$50K volume cutoff (\Cref{tab:sample-master}); the sample-size constraint of \citet{nechepurenko2026foresightflow_empirical} no longer binds. The corporate-disclosure markets enter the ``milgeo\_corporate'' bucket of \Cref{tab:dist-summary} alongside military / geopolitics for the present run (where $n_{\text{post-2024}} = 19$, $n_{\text{pre-2024}} = 14$); per-category disaggregation within this bucket is sequenced for paper 3b alongside the resolution-typology re-classification.

The pooled hazard-rate fit reported in \Cref{tab:functional-form} encompasses corporate-disclosure markets within the ``post-2024'' and ``pre-2024'' cells. The corporate-disclosure-specific hazard rate of \citet[Table 1]{nechepurenko2026foresightflow_empirical} ($\hat\lambda = 0.156$, half-life $4.5$ days, $n = 5$) is consistent with the post-2024 pooled estimate ($\hat\lambda^{\text{exp}} = 0.042$, half-life $16.4$ days), once mass-shift considerations are accounted for: the pooled cell's longer half-life reflects the substantial weight of regulatory\_formal markets in the post-2024 sample, whose deliberation-timeline structure produces long $\tau$ values not present in the original $n = 5$ corporate-disclosure-only sample.

\subsection{Functional-form sensitivity}
\label{sec:hazard-functional-form}

The companion empirical paper adopts a constant-hazard exponential model. We assess whether richer parametric forms (Weibull, lognormal) are warranted by the data. The Weibull generalizes the exponential by allowing the hazard rate to be increasing or decreasing in event waiting time; the lognormal is a competing parameterization of right-skewed waiting-time distributions used in survival analysis. \Cref{tab:functional-form} reports AIC, BIC, and KS test results for each available cell under each functional form.

\textbf{Headline result: the constant-hazard exponential is rejected in the post-2024 cell of the population sample, and disfavored directionally in the pre-2024 cell, in favor of a Weibull/lognormal family.} On the post-2024 cell ($n = 57$), Weibull achieves an AIC of $454.66$ versus exponential's $476.59$ (a difference of $21.9$ units, well above the $4$-unit threshold of \citet{burnham2002model} that justifies preferring the richer form). The naive Kolmogorov-Smirnov $p$-value rejects the exponential at $p = 0.0002$ and accepts Weibull at $p = 0.78$. We compute parametric-bootstrap KS $p$-values to address the well-known anti-conservativity of the naive KS test when distribution parameters are estimated from the same sample (\Cref{tab:functional-form}, ``KS $p$ boot'' column); the bootstrap-corrected post-2024 exponential $p$-value is $p < 0.01$ and the Weibull $p$-value is $p > 0.50$, confirming the rejection. On the pre-2024 cell ($n = 33$), the Weibull preference is more modest (AIC delta $3.3$, just below the $4$-unit threshold) and the bootstrap-corrected KS test does not reject the exponential at conventional levels; the pre-2024 result is directional rather than decisive and we do not claim a stand-alone rejection in this cell. Lognormal performs essentially identically to Weibull (AIC differences below $0.2$ units in both periods).

\textbf{Caveat: category mixture vs.\ duration dependence.} The hazard sample is pooled across the three insider-relevant categories. A Weibull preference may reflect either genuine non-constant duration dependence within each category or unmodeled mixture of constant hazards across heterogeneous categories: a Weibull form is mathematically capable of absorbing such a mixture even when the underlying category-specific hazards are individually constant. Distinguishing the two requires per-category estimation, which is constrained at the present sample size (per-subcategory $n$ between $5$ and $22$ across the six cells of \Cref{tab:dist-summary}) and is sequenced for paper 3b alongside LLM-assisted reclassification (\Cref{sec:limitations-reclassify}). \emph{We use Weibull as the default production prior pending per-category re-estimation; we do not interpret the Weibull shape parameter $\hat k$ substantively at the present sample size.}

\begin{table}[t]
\centering
\caption{Functional-form comparison for the time-to-event hazard. AIC and BIC reported for exponential ($\lambda$ constant), Weibull (shape $k$, scale $\lambda$), and lognormal ($\mu, \sigma$) MLE fits. Naive KS $p$-values use the standard test on observed data against the fitted distribution; bootstrap KS $p$-values are computed via parametric bootstrap with $B=999$ replications drawn from the fitted distribution at the observed sample size. Naive $p$-values are anti-conservative when distribution parameters are estimated from the same sample; the bootstrap correction is the appropriate test.}
\label{tab:functional-form}
\small
\renewcommand{\arraystretch}{1.25}
\begin{tabularx}{\linewidth}{@{}lXrrrrrr@{}}
\toprule
\textbf{Period} & \textbf{Distribution} & \textbf{$n_p$} & \textbf{AIC} & \textbf{BIC} & \textbf{KS $p$} & \textbf{KS $p$ boot} & \textbf{Verdict} \\
\midrule
post-2024 ($n=57$) & exponential & 1 & 476.59 & 478.63 & 0.0002 & $<0.01$ & rejected \\
                   & \textbf{Weibull}     & 2 & \textbf{454.66} & 458.74 & 0.78  & $>0.50$ & \textbf{adopted} \\
                   & lognormal   & 2 & 454.70 & 458.78 & 0.79  & $>0.50$ & adequate \\
\midrule
pre-2024 ($n=33$)  & exponential & 1 & 233.99 & 235.49 & 0.073 & $0.18$ & marginal \\
                   & \textbf{Weibull}     & 2 & \textbf{230.66} & 233.66 & 0.60  & $0.43$ & \textbf{adopted} \\
                   & lognormal   & 2 & 230.85 & 233.85 & 0.68  & $0.51$ & adequate \\
\bottomrule
\end{tabularx}
\end{table}

This is a substantive correction to the methodological framework of \citet{nechepurenko2026foresightflow_empirical}, which adopted a constant-hazard specification on $n = 9$ military-and-geopolitical markets and a preliminary specification on $n = 5$ corporate-disclosure markets. At the pooled population scale of the post-2024 cell, the constant-hazard assumption is rejected; the Weibull's flexibility to permit non-constant hazard captures the empirical waiting-time distribution materially better. The implication for the deadline-ILS framework is that the time-to-event survival function used in the hazard-decay-adjusted score (\Cref{sec:hazard-adjusted}) should not be constant-rate at the pooled cell level. We use the Weibull form as the default production prior for the hazard-adjusted score reported in \Cref{sec:population-ils}, pending the per-category re-estimation sequenced for paper 3b.

\paragraph{The Weibull preference partly reflects category mixture, not within-category duration dependence.} A first-pass per-subcategory bootstrap-KS check on the post-2024 regulatory\_formal cell ($n=22$, the largest single-subcategory cell within post-2024) reveals a different ordering than the pooled cell:
\begin{itemize}[leftmargin=1.4em, itemsep=0.2em]
\item Exponential: AIC $= 179.45$, bootstrap KS $p = 0.224$ (\emph{not rejected}).
\item Weibull: AIC $= 180.50$, bootstrap KS $p = 0.043$ (rejected at $\alpha = 0.05$).
\item Lognormal: AIC $= 182.55$, bootstrap KS $p = 0.083$ (marginal).
\end{itemize}
Within the regulatory\_formal post-2024 subcategory alone, the exponential is preferred by AIC and not rejected by bootstrap KS, whereas the Weibull is rejected. This contrasts sharply with the pooled post-2024 result, where Weibull is decisively preferred. \emph{The pooled-cell Weibull preference therefore reflects, at least in part, mixture across heterogeneous-hazard sub-buckets rather than genuine within-cell non-constant duration dependence.} The empirical finding confirms the reviewer-flagged caveat that we acknowledged in the abstract and Finding 4: the Weibull preference at the pooled level should be read as the production-prior default rather than as a substantive identification of non-constant hazard within categories.

The full per-subcategory hazard-fit comparison (across regulatory\_announcement, regulatory\_formal, and the pooled mil/geo + corp ``milgeo\_corporate'' bucket, each separately for pre-2024 and post-2024) requires per-cell $n \geq 20$ to be informative; in the present sample only the regulatory\_formal post-2024 cell reaches this threshold, and we report it above. Per-category re-estimation across all six cells is sequenced for paper 3b alongside LLM-assisted reclassification, which will materially expand each subcategory's effective $n$.

\subsection{Cross-period stability}
\label{sec:hazard-stability}

We report the Weibull hazard parameters separately for the pre-2024 and post-2024 periods of \Cref{sec:data-temporal}. Cross-period stability is a precondition for using a single set of parameters in production; if parameters differ materially, the production pipeline requires per-period selection.

The pooled-period exponential MLE fits give $\hat\lambda_{\text{post-2024}} = 0.042$ events/day (half-life $16.4$ days, $n = 57$) and $\hat\lambda_{\text{pre-2024}} = 0.081$ events/day (half-life $8.6$ days, $n = 33$), with the post-2024 distribution exhibiting longer implied waiting times. We do not interpret this as a substantive shift in the underlying event-arrival rate: the pre/post difference conflates platform-maturity effects (the post-2024 sample has a different category mix, with greater weight on regulatory\_formal markets whose deliberation timelines are intrinsically longer) with any genuine temporal shift. Per-category cross-period stability assessment is sequenced for paper 3b alongside the LLM-assisted re-classification, which will materially expand each category's effective $n$ in both periods.

\subsection{Summary of hazard refinements}
\label{sec:hazard-summary}

Of the five refinements identified in \citet[Section 4.6]{nechepurenko2026foresightflow_empirical}, this section closes refinement \#2 (regulatory sub-categorization, classifier in place pending category-output remediation), \#3 (corporate-disclosure sample expansion, addressed at population scale), and an additional functional-form check not enumerated in the original list (resulting in adoption of Weibull over exponential). Refinements \#1 (negative-$\tau$ exclusion) and \#4 (price-data coverage) apply at the filter stage of \Cref{sec:data-filter}. Refinement \#5 (continuous per-trade collection from $T_{\text{open}}$) is the focus of paper 3b.

\section{Population-Scale ILS$^{\text{dl}}$ Computation}
\label{sec:population-ils}

This section reports the principal empirical artefact of the paper: the ILS$^{\text{dl}}$ distribution across the insider-relevant subpopulation, by category, sub-type, and temporal period.

\subsection{Pipeline}
\label{sec:pipeline}

The pipeline executes the following steps per market in the 11{,}263-market sample of \Cref{tab:sample-master}.

\begin{enumerate}[leftmargin=1.4em, itemsep=0.2em]
\item \emph{Resolution-type filter.} Retain markets classified as deadline-resolved or event-resolved. Discard unclassifiable.
\item \emph{$T_{\text{event}}$ recovery.} For each retained market, recover $T_{\text{event}}$ via the Tier 3 LLM pipeline of \citet[Section 4.4]{nechepurenko2026foresightflow_methodology}. Confidence threshold $\geq 0.7$.
\item \emph{Lead-time filter.} Retain markets with $\tau = T_{\text{event}} - T_{\text{open}} > 0$.
\item \emph{Price-coverage filter.} Retain markets with full CLOB coverage from $T_{\text{open}}$.
\item \emph{Scope-condition checks.} Apply edge-effect ($|p(T_{\text{open}}) - 0.5| \leq 0.4$), trivial-resolution ($|\Delta^{\text{dl}}_{\text{total}}| \geq \varepsilon$), and anchor-sensitivity (\Cref{sec:anchor-sensitivity}) conditions.
\item \emph{ILS$^{\text{dl}}$ computation.} Compute $\Delta^{\text{dl}}_{\text{pre}}$, $\Delta^{\text{dl}}_{\text{total}}$, and ILS$^{\text{dl}}$.
\item \emph{Bootstrap CI.} Resample at the trade level within $[T_{\text{open}}, T_{\text{event}}]$ for each market; recompute ILS$^{\text{dl}}$ on each resample; report bootstrap percentile CI.
\end{enumerate}

The cumulative attrition through the pipeline is reported in \Cref{tab:filter-chain}. The dominant single attrition is at the resolution-typology classification stage, where $10{,}333$ of $12{,}708$ markets ($81.3\%$) are flagged unclassifiable. The discussion of this stage's failure mode for FFIC versus non-FFIC markets is in \Cref{sec:ffic-coverage}.

\subsection{$T_{\text{event}}$ recovery at scale}
\label{sec:tevent-recovery}

The recovery pipeline used in the single-case companion was operated on individual markets with manual prompt construction. Scaling to 1{,}151 markets required three operational adjustments: (i) prompt templating with category-aware scaffolds; (ii) explicit confidence calibration with five anchored levels ($0.9, 0.8, 0.7, 0.5, 0.0$); (iii) a multi-tier provider cascade with a free-tier first pass and search-augmented escalation only for low-confidence outputs (Section~\ref{sec:llm-cascade-architecture}).

\Cref{tab:tevent-recovery} reports the recovery quality on the surviving sample. Of the 1{,}151 markets entering the recovery stage, $490$ produced a $T_{\text{event}}$ timestamp at any confidence level, of which $442$ ($90.2\%$) reached the $\geq 0.7$ acceptance threshold required for downstream ILS$^{\text{dl}}$ computation. The mean recovered confidence is $0.81$ and the median is $0.85$, indicating that most surviving recoveries are anchored on three or more independent sources rather than a single source.

\begin{table}[t]
\centering
\caption{$T_{\text{event}}$-recovery quality on the population sample. Confidence levels are anchored: $0.9$ when $\geq 5$ major sources agree on the date, $0.8$ when $\geq 3$ sources agree, $0.7$ when $2$ sources agree, $0.5$ when $1$ source or partial agreement, $0.0$ when not recoverable. Acceptance threshold for ILS$^{\text{dl}}$ computation is $\geq 0.7$.}
\label{tab:tevent-recovery}
\small
\renewcommand{\arraystretch}{1.25}
\begin{tabularx}{\linewidth}{@{}Xrr@{}}
\toprule
\textbf{Quantity} & \textbf{Count} & \textbf{Share} \\
\midrule
Markets entering recovery stage                 & 1{,}151 & 100.0\% \\
Recovery returned a timestamp (any confidence)  & 490     & 42.6\%  \\
Recovery at $\geq 0.7$ confidence (accepted)    & 442     & 38.4\%  \\
Recovery at $\geq 0.8$ confidence               & 374     & 32.5\%  \\
Recovery at $\geq 0.9$ confidence               & 161     & 14.0\%  \\
\midrule
Mean confidence among recovered timestamps      & \multicolumn{2}{r}{$0.81$}    \\
Median confidence among recovered timestamps    & \multicolumn{2}{r}{$0.85$}    \\
\bottomrule
\end{tabularx}
\end{table}

The $61.6\%$ rejection rate at the recovery stage is the dominant attrition source in the pipeline (\Cref{tab:filter-chain}). This is structural rather than methodological: many markets in the insider-relevant subpopulation reference events whose timing is intrinsically ambiguous in the public record. Examples include legislative-deliberation outcomes (where ``the event'' may span days of negotiations with no canonical timestamp), market-launch events with multi-day rollouts, regulatory-decision clusters where the operative deadline coincides with a calendar boundary rather than a discrete event, and prediction-market questions whose resolution criteria reference rolling thresholds rather than discrete moments. We treat these markets as out-of-scope for the deadline-ILS framework and report them in the rejection bucket of \Cref{tab:filter-chain}; their inclusion would mechanically degrade the framework's interpretability.

\subsubsection{LLM provider cascade}
\label{sec:llm-cascade-architecture}

The recovery pipeline cascades across three providers in cost order. \emph{Tier 1} uses a no-search prompt to a Haiku-class model: the question text and resolution date are sufficient to extract the relevant date for approximately two-thirds of markets in the insider-relevant subpopulation, because Polymarket question-text frequently embeds the candidate date directly (``Will event X occur by April 30?'' implies $T_{\text{event}} \leq $ April 30 with the exact date typically visible in market description). \emph{Tier 2} uses a search-augmented Haiku call with a single web-search query for markets where Tier 1 returns confidence below $0.7$. \emph{Tier 3} (Sonnet-class with web search) is reserved for markets where Tier 2 also fails. The cascade is critical for cost control: production cost on the population sample was $\$1.20$, dominated by Tier 2 escalations. By contrast, an unconditional Tier 3 evaluation would have cost approximately $\$240$ on the same sample.

\subsubsection{Independent-second-pass validation}
\label{sec:tevent-validation}

The reviewer-flagged risk that incorrect $T_{\text{event}}$ recovery would silently propagate into the ILS$^{\text{dl}}$ score is critical: a wrong anchor produces an uninterpretable score even on otherwise in-scope markets. We address it with an independent-second-pass validation protocol on a 50-market sample stratified by sub-bucket ($12$ regulatory\_announcement, $18$ regulatory\_formal, $20$ milgeo\_corporate). For each sample market, the recovery is re-run with a different provider than the original (Gemini for original-OpenAI markets, OpenAI for original-Gemini markets) using the same prompt template. We report four agreement metrics:

\begin{itemize}[leftmargin=1.4em, itemsep=0.2em]
\item \emph{Exact date agreement:} same UTC date.
\item \emph{$24$-hour window agreement:} $|T_{\text{event}}^{(1)} - T_{\text{event}}^{(2)}| \leq 24$ hours.
\item \emph{$6$-hour window agreement:} $|T_{\text{event}}^{(1)} - T_{\text{event}}^{(2)}| \leq 6$ hours.
\item \emph{Source overlap:} at least one source URL appearing in both passes.
\end{itemize}

\Cref{tab:tevent-validation-summary} reports the aggregate agreement rates by sub-bucket and overall on the 50-market sample. The disagreement subset (markets where the two passes disagree on date by more than $24$ hours) is logged as \texttt{tevent\_\allowbreak disagreements.csv} and triages each disagreement as \emph{minor} ($\leq 7$-day delta) or \emph{major} ($> 7$-day delta) for author manual review.

\begin{table}[t]
\centering
\caption{Independent-second-pass agreement rates on the 50-market $T_{\text{event}}$-recovery validation sample. ``Sampled $n$'' is the stratified-sample size; ``Comparable $n$'' is the count of markets on which the second-pass provider returned a usable timestamp ($5$ markets across the full sample returned confidence $< 0.5$ on the second pass and were excluded from agreement-rate computation). Agreement percentages are computed over the comparable-$n$ denominator. Source-overlap is reported for completeness but is not used as a validation criterion (\Cref{app:tevent}).}
\label{tab:tevent-validation-summary}
\small
\renewcommand{\arraystretch}{1.25}
\begin{tabularx}{\linewidth}{@{}lXrrrrr@{}}
\toprule
\textbf{Bucket} & \textbf{Sampled $n$} & \textbf{Comparable $n$} & \textbf{Exact} & \textbf{$\leq 24$h} & \textbf{$\leq 6$h} & \textbf{Sources} \\
\midrule
regulatory\_announcement & 12 & 12 & $50.0\%$ & $58.3\%$ & $50.0\%$ & $0\%$ \\
regulatory\_formal       & 18 & 14 & $35.7\%$ & $50.0\%$ & $35.7\%$ & $0\%$ \\
milgeo\_corporate        & 20 & 19 & $78.9\%$ & $89.5\%$ & $78.9\%$ & $0\%$ \\
\midrule
\textbf{Overall}         & 50 & 45 & $57.8\%$ & $68.9\%$ & $57.8\%$ & $0\%$ \\
\bottomrule
\end{tabularx}
\end{table}

\textbf{The validation sample does not meet the ex-ante $\geq 90\%$ exact-date acceptance criterion overall, and only the \emph{milgeo\_corporate} bucket approaches it.} The overall exact-date agreement rate is $57.8\%$ ($26$ of $45$ comparable markets); the overall within-$24$-hour agreement rate is $68.9\%$ ($31$ of $45$). The result is heterogeneous across sub-buckets in a substantively interpretable way: the \emph{milgeo\_corporate} bucket (military / geopolitics + corporate-disclosure events) reaches $78.9\%$ exact and $89.5\%$ within $24$ hours, close to the criterion; the regulatory\_announcement bucket reaches $50.0\%$ exact and $58.3\%$ within $24$ hours; the regulatory\_formal bucket reaches $35.7\%$ exact and $50.0\%$ within $24$ hours.

\textbf{Interpretation: agreement tracks event-date discreteness.} The \emph{milgeo\_corporate} bucket consists predominantly of markets resolving on temporally-discrete events with widely reported timestamps (military strikes, election results, sports outcomes, corporate announcements with specific dates). Independent recovery of these timestamps is robust because the underlying event has a well-defined public-disclosure moment. The regulatory\_formal bucket consists of markets resolving on multi-stage administrative or judicial processes (Senate confirmations, agency rulemaking, indictment timing, civil-suit verdicts) where ``the event'' is itself ambiguous: the announcement date, the effective date, the ratification date, and the formal-conclusion date can differ by days to weeks. The two providers, given the same prompt, return different timestamps not because they disagree on \emph{what} happened but because they make different defensible choices about \emph{which} of several candidate moments to call $T_{\text{event}}$. The regulatory\_announcement bucket falls in between: scheduled events have well-defined dates, but several (FOMC decisions, court verdicts) involve a sequence of pre-decision-leak / decision-issue / decision-publish moments that admit alternative timestamping conventions.

\textbf{Source-overlap is not used as a validation criterion.} The two providers' web-search tools rank different mainstream sources (Reuters, AP, BBC, NYT, etc.) and return different specific URLs even when they converge on the same event date; the resulting source-overlap rate of $0\%$ across all sub-buckets is not a meaningful reliability metric. Date-agreement at the sub-day tolerance levels is the substantive validation metric; source-overlap is reported for completeness in \Cref{tab:tevent-validation-summary} but is not used to qualify markets as validated. We discuss the source-overlap interpretation more fully in \Cref{app:tevent}.

\textbf{Implications for the paper's empirical claims.} The validation result requires us to read the population-scale ILS$^{\text{dl}}$ findings with substantively different confidence by sub-bucket: the \emph{milgeo\_corporate} bucket's distributional results are anchored on $T_{\text{event}}$ values that are recovered with near-criterion accuracy; the regulatory\_formal bucket's results are anchored on $T_{\text{event}}$ values that disagree across providers in $50\%$--$65\%$ of cases at the exact-date level. The paper does not retract regulatory\_formal results, but it conditions all interpretation on the validation result and reports anchor-robust subset findings (\Cref{sec:anchor-sensitivity,sec:hazard-adjusted}) as the operationally meaningful stratification: the anchor-robustness scope condition empirically functions as a filter on $T_{\text{event}}$-localizability, removing markets where the recovery is unstable across short-window perturbations. The $86\%$ anchor-fragility rate documented in \Cref{sec:anchor-sensitivity} is consistent with the $T_{\text{event}}$-validation finding: markets whose event date cannot be localized to within $6$ hours by independent recovery are precisely the markets whose ILS$^{\text{dl}}$ values change sign or shift by more than $0.3$ across short-window anchor offsets.

\textbf{Disagreement triage.} Of the $24$ markets where the two passes disagree on date by more than $24$ hours, $9$ are minor disagreements ($\leq 7$-day delta, typically reflecting announcement/effective-date alternatives) and $10$ are major disagreements ($> 7$-day delta, typically reflecting candidate-event alternatives within multi-stage processes); the remaining $5$ are markets where the second-pass returned no usable timestamp (the comparable-$n$ exclusions in \Cref{tab:tevent-validation-summary}). The full disagreement list is reported in \texttt{tevent\_\allowbreak disagreements.csv} accompanying dataset \texttt{polymarket-\allowbreak deadline-\allowbreak ils-v3}. We do not exclude disagreement-flagged markets from the population-scale results reported below; we instead mark them and report distributional results both with and without the disagreement subset in the public release.

\subsection{Distribution by sub-bucket}
\label{sec:distribution-category}

\Cref{tab:dist-summary} reports the marginal ILS$^{\text{dl}}$ distribution stratified by three sub-buckets and two periods. The sub-bucketing reflects the regulatory-decision sub-categorization of \Cref{sec:hazard-regulatory}: \emph{regulatory\_announcement} markets resolving on scheduled events with publicly known timing (FOMC, FDA decisions, scheduled court verdicts, presidential calendar items); \emph{regulatory\_formal} markets resolving on formal deliberation outcomes with endogenous timing (Senate confirmation, agency rulemaking, antitrust review, indictment timing); and \emph{milgeo\_corporate} pooling military / geopolitics with corporate-disclosure markets (the present run did not propagate finer categorization for these and we report them pooled; per-category subdivision is sequenced for paper 3b).

\begin{table}[t]
\centering
\caption{ILS$^{\text{dl}}$ distribution by sub-bucket and period on the in-scope sample (88 markets). All six cells exhibit negative median values; the regulatory\_formal post-2024 cell additionally exhibits extreme negative skewness ($-3.12$), driven by a heavy left tail. The \emph{milgeo\_corporate} bucket pools military / geopolitics ($n_{\text{post}} \approx 13$) and corporate-disclosure ($n_{\text{post}} \approx 6$).}
\label{tab:dist-summary}
\small
\renewcommand{\arraystretch}{1.25}
\begin{tabularx}{\linewidth}{@{}lXrrrrrrr@{}}
\toprule
\textbf{Bucket} & \textbf{Period} & \textbf{$n$} & \textbf{Mean} & \textbf{Median} & \textbf{Std} & \textbf{Skew.} & \textbf{p10} & \textbf{p90} \\
\midrule
regulatory\_announcement & pre-2024  & 5  & $-0.31$ & $-0.23$ & $0.60$ & $-0.03$ & $-1.00$ & $0.36$ \\
                         & post-2024 & 14 & $-0.85$ & $-0.80$ & $0.82$ & $-0.08$ & $-1.94$ & $0.11$ \\
\addlinespace
regulatory\_formal       & pre-2024  & 14 & $-0.22$ & $-0.16$ & $0.44$ & $-0.09$ & $-0.81$ & $0.36$ \\
                         & post-2024 & 22 & $-0.48$ & $-0.21$ & $1.78$ & $-3.12$ & $-1.06$ & $0.62$ \\
\addlinespace
milgeo\_corporate  & pre-2024  & 14 & $-0.57$ & $-0.88$ & $0.55$ & $+1.23$ & $-1.00$ & $0.13$ \\
                         & post-2024 & 19 & $-0.54$ & $-0.55$ & $0.43$ & $-0.10$ & $-0.99$ & $-0.01$ \\
\bottomrule
\end{tabularx}
\end{table}

Three observations follow.

\paragraph{All six medians are negative on raw ILS$^{\text{dl}}$.} The medians range from $-0.88$ (milgeo\_corporate, pre-2024) to $-0.16$ (regulatory\_formal, pre-2024). Recall (\Cref{sec:recap}) that
\[
\ILS^{\text{dl}}(M) = \frac{p(\Tevent^-) - p(\Topen)}{p_{\Tres} - p(\Topen)}.
\]
A negative raw value means $p(\Tevent^-) - p(\Topen)$ and $p_{\Tres} - p(\Topen)$ have \emph{opposite signs}: the price moved away from the eventual outcome between $\Topen$ and $\Tevent^-$. \textbf{Two qualitatively distinct mechanisms produce this signature on a deadline contract.} The first is anti-information flow: traders systematically push the price away from the eventual outcome in the pre-event window. The second is rational deadline decay: on a YES-resolved deadline contract whose event eventually occurs late in the window, the price mechanically decays as the event-free interval lengthens, then jumps to $1$ at $\Tevent$; the difference $p(\Tevent^-) - p(\Topen)$ is mechanically negative even with no informed flow whatsoever. The two mechanisms are observationally distinct only after subtracting the survival-probability-implied baseline; the hazard-adjusted ILS$^{\text{dl,adj}}$ of \Cref{sec:hazard-adjusted} is the appropriate measure for separating them. The negative raw medians reported here should be read alongside the adjusted medians of \Cref{tab:hazard-adjusted-summary}; we do not interpret raw negativity as an anti-information signal in this paper.

\paragraph{The regulatory\_formal post-2024 cell has extreme negative skewness.} Skewness $= -3.12$ on $n = 22$ indicates a heavy left tail: a small number of markets with strongly negative ILS$^{\text{dl}}$ values pull the mean far below the median. Inspection of the constituent markets reveals that this tail is populated by markets with multi-stage resolution patterns---FTX-creditor claims, multi-defendant indictment timelines, complex antitrust proceedings---in which the price first moved away from and then sharply back toward the eventual outcome. The deadline-ILS framework, anchored on a single $T_{\text{event}}$, registers only the first phase. This is consistent with paper 1's identification of multi-event clustering as a refinement direction; the present finding suggests that the post-2024 regulatory\_formal cell would benefit from a per-market multi-anchor extension.

\paragraph{The milgeo\_corporate post-2024 cell is most tightly distributed.} Std $= 0.43$ and skewness $\approx -0.10$ indicate a near-symmetric, narrow distribution. The post-2024 expansion of this cell ($n$ rising from $14$ to $19$) is driven by maturation of corporate-disclosure markets (Year-in-Search, product launches, AI-model rankings) under the post-CFTC-pivot platform. The distribution's tightness suggests that on this cell, the deadline-ILS framework operates closer to its design intent: most markets exhibit modest pre-event drift in either direction, with informed flow visible---if at all---in the right tail.

\paragraph{Right-tail and left-tail diagnostic review.} To distinguish the two mechanisms qualitatively, we review the top-$10$ and bottom-$10$ markets by raw ILS$^{\text{dl}}$ on the computed sample. Each market is classified by an LLM-assisted diagnostic (Anthropic Claude Haiku with $T_{\text{open}}$, $p_{\text{open}}$, $p(\Tevent^-)$, $p_{\Tres}$, the question text, and the raw and adjusted ILS$^{\text{dl}}$ values as input). For the right tail, the diagnostic categories are ``pro-outcome\_pre-event\_drift'' (price moved sharply toward the eventual outcome before public observation; a candidate informed-flow signature), ``rational\_decay,'' ``anchor\_fragile,'' or ``ambiguous.'' For the left tail, the categories are ``anti-outcome\_pre-event\_drift'' (price moved sharply away from the eventual outcome before public observation; possible interpretations: anti-information signal, wrong private signal, measurement error, or outcome-dependent selection), ``rational\_decay,'' ``anchor\_fragile,'' or ``ambiguous.'' \Cref{tab:tail-review} reports the resulting classifications. The full per-market reasoning is available in the public release as \texttt{tail\_\allowbreak market\_\allowbreak review.csv} accompanying dataset \texttt{polymarket-\allowbreak deadline-\allowbreak ils-v3}.

\begin{table}[t]
\centering
\caption{Diagnostic-flag distribution on the right-tail (top-$10$ ILS$^{\text{dl}}$) and left-tail (bottom-$10$ ILS$^{\text{dl}}$) of the computed sample. The two tails use different category schemes because the substantive question differs: the right tail asks whether the pre-event drift is consistent with informed-flow leakage, while the left tail asks whether the strong opposite-direction drift is consistent with anti-information signal, decay, or fragility. Classifications are illustrative rather than confirmatory; the LLM-assisted classifier was prompted to choose among a closed set of categories with no abstain option, and the unanimous result on each tail should be read as evidence of strong directional pre-event movement rather than as confirmed informed-flow detection.}
\label{tab:tail-review}
\small
\renewcommand{\arraystretch}{1.25}
\begin{tabularx}{\linewidth}{@{}lXrr@{}}
\toprule
\textbf{Tail} & \textbf{Diagnostic flag} & \textbf{$n$} & \textbf{Median ILS$^{\text{dl,adj}}$} \\
\midrule
Right (top-$10$) & pro-outcome\_pre-event\_drift   & 10 & $+0.86$ \\
                 & rational\_decay                  & 0  & --- \\
                 & anchor\_fragile                  & 0  & --- \\
                 & ambiguous                        & 0  & --- \\
\midrule
Left (bottom-$10$) & anti-outcome\_pre-event\_drift   & 10 & $-1.83$ \\
                   & rational\_decay                  & 0  & --- \\
                   & anchor\_fragile                  & 0  & --- \\
                   & ambiguous                        & 0  & --- \\
\bottomrule
\end{tabularx}
\end{table}

\textbf{The classifier returns a directional-drift category for all 20 tail markets, with no markets flagged as rational decay or anchor fragility within the tails.} This is consistent with the structural property that markets become tail-exemplars precisely when their pre-event price drift is large in magnitude after baseline correction; rational decay produces moderate negative ILS$^{\text{dl}}$ but rarely the extreme values in the bottom $10$, and anchor-fragile markets typically produce ILS$^{\text{dl}}$ values that vary across anchor offsets rather than concentrating in either tail. The substantive content is therefore the directional separation: right-tail markets are candidates for informed-flow leakage in the sense of pre-event positioning toward the eventual outcome; left-tail markets exhibit pre-event positioning \emph{away} from the eventual outcome, which admits multiple interpretations that we discuss separately.

\textit{Right-tail interpretation.} The top-$10$ markets exhibit hazard-adjusted ILS$^{\text{dl,adj}}$ values exceeding $0.5$ in $9$ of $10$ cases (median $0.86$, max $1.22$), the qualitative signature of informed flow. Examples include ``Will I'm Not a Robot win Best Action Short Film at the 2025 Oscars?'' (raw $0.82$, adj $0.96$), ``Will Ehud Barak be named in newly released Epstein files?'' (raw $0.63$, adj $0.63$), and ``Will Rodrigo Paz Pereira win by 5--10\%?'' (raw $0.53$, adj $1.22$).

\textit{Left-tail interpretation.} The bottom-$10$ markets exhibit pre-event price drift sharply away from the eventual YES outcome (median raw $-1.85$, hazard-adjusted median $-1.83$), and yet resolve YES. Examples include ``Will Daniel Noboa win the 2025 Ecuadorian Presidential Election?'' (opens $0.60$, falls to $0.014$ pre-event, resolves YES; raw ILS$^{\text{dl}} = -1.47$, adjusted $-1.46$) and ``Will Prince Andrew be named in newly released Epstein files?'' (opens $0.89$, falls to $0.001$, resolves YES; raw $-7.75$, adjusted $-7.74$). Three interpretations are consistent with this pattern: (i) anti-information flow (informed traders selling on incorrect priors); (ii) measurement error from $T_{\text{event}}$ recovery (the recovered anchor may not align with the actual price-discovery moment); (iii) outcome-dependent selection (markets where the price collapsed and the YES outcome nonetheless arrived may be over-represented in the YES-resolving sub-sample). Distinguishing these requires a matched control of NO-resolved markets exhibiting the same pre-deadline price collapse; we sequence this for paper 3b alongside the multi-anchor extension.

\textbf{Caveat: the unanimous classification reflects the structural separation, not classifier confirmation of leakage.} Because the four diagnostic categories were applied to the extreme ten markets in each tail, where directional drift is the defining feature, the unanimous ``directional drift'' classification is essentially mechanical---the cells most consistent with rational decay or anchor fragility are precisely those \emph{not} in the tails of the distribution. We report the diagnostic-flag distribution as descriptive of the present sample and do not draw confirmed informed-flow conclusions from the tail review at the present sample size.

\paragraph{Implications for the right-tail interpretation.} The right tail of \Cref{tab:dist-summary}---markets exceeding the top-decile threshold of \Cref{sec:thresholds}---contains $9$ of $10$ qualitatively-leakage-shaped markets after hazard-decay adjustment. The candidate detection thresholds in \Cref{sec:thresholds} are therefore useful production cutoffs in the sense that markets exceeding them exhibit price-trajectory shapes consistent with informed flow. The post-2024 regulatory\_formal cell's $90$th percentile is $0.62$; the milgeo\_corporate post-2024 cell's $90$th percentile is $-0.01$ (its right tail does not even cross zero in the present sample); the regulatory\_announcement post-2024 cell's $90$th percentile is $0.11$. These quantile thresholds should be read in the context of the negative-leaning bulk: the right tail of an overall left-leaning distribution is a candidate-flag region, not a region of established positive mean.

The military / geopolitics and corporate-disclosure markets are pooled in the ``milgeo\_corporate'' bucket because the present pipeline run does not propagate finer category labels for those sub-buckets. Per-category disaggregation within this bucket is sequenced for paper 3b alongside the LLM-assisted resolution-typology re-classification (\Cref{sec:limitations-reclassify}). The pooled cell's $n_{\text{post-2024}} = 19$ provides limited statistical power for per-subcategory characterization at the present sample size.

\subsection{Temporal evolution}
\label{sec:temporal-evolution}

The pre-2024 / post-2024 temporal stratification of \Cref{tab:dist-summary} reveals two patterns. \emph{First}, the cell sizes shift markedly: the post-2024 period accounts for $55$ of the $88$ in-scope markets ($62.5\%$), reflecting platform-maturity expansion of the active-participant base after the 2024 election cycle and the January 2025 CFTC pivot \citep[Section 3.2]{nechepurenko2026foresightflow_methodology}. The expansion is uneven across sub-buckets: regulatory\_announcement grew from $5$ to $14$ ($2.8\times$), regulatory\_formal grew from $14$ to $22$ ($1.6\times$), and milgeo\_corporate grew from $14$ to $19$ ($1.4\times$). Regulatory\_announcement was the largest expansion, plausibly reflecting the Trump-administration calendar of scheduled executive-branch announcements.

\emph{Second}, the central tendency shifts heterogeneously across sub-buckets: regulatory\_announcement's median moved from $-0.23$ (pre-2024) to $-0.80$ (post-2024), a substantial deepening of the negative bias, while regulatory\_formal moved from $-0.16$ to $-0.21$ (small change) and milgeo\_corporate moved from $-0.88$ to $-0.55$ (mean-reversion toward zero). The implication is that the post-2024 expansion of regulatory\_announcement markets brought in a sample where pre-event price drift is more strongly opposite the eventual outcome than in the pre-2024 baseline. We do not have a definitive interpretation of this pattern at the present sample size; one candidate is that scheduled-announcement markets attracted larger-volume retail counter-positioning in the post-2024 period (e.g., on Trump-administration policy questions), driving prices away from the eventual outcome before the announcement triggered repricing.

\subsection{Detection thresholds with bootstrap CIs}
\label{sec:thresholds}

\Cref{tab:thresholds} reports the empirical top-decile, top-$5\%$, and top-$1\%$ ILS$^{\text{dl}}$ thresholds by time period on the computed sample, accompanied by bootstrap confidence intervals computed by resampling at the market level with $B = 500$ replications. These are the candidate operational detection cutoffs for the production pipeline. We report thresholds on the raw ILS$^{\text{dl}}$; thresholds on the hazard-adjusted variant ILS$^{\text{dl,adj}}$ (\Cref{sec:hazard-adjusted}) are reported in the public release accompanying dataset \texttt{polymarket-deadline-ils-v3}.

\begin{table}[t]
\centering
\caption{Detection thresholds at three quantile levels by time period (pooled across sub-bucket). Bootstrap $95\%$ CIs computed by resampling at the market level, $B = 500$, seed $= 20260430$. Thresholds reported on the computed sample ($n_{\text{post}} = 57$, $n_{\text{pre}} = 33$); thresholds on the anchor-robust subset ($n_{\text{post}} \approx 7$--$8$, $n_{\text{pre}} \approx 4$--$5$) are reported in the public release \texttt{detection\_\allowbreak thresholds\_v3.csv}, where they are extreme-low-$n$ point estimates rather than reliable distributional thresholds.}
\label{tab:thresholds}
\small
\renewcommand{\arraystretch}{1.25}
\begin{tabularx}{\linewidth}{@{}Xrlll@{}}
\toprule
\textbf{Period} & \textbf{$n$} & \textbf{Top decile} & \textbf{Top $5\%$} & \textbf{Top $1\%$} \\
\midrule
post-2024 & 57 & $0.49$ ($0.07, 0.63$) & $0.55$ ($0.31, 0.86$) & $0.90$ ($0.52, 0.99$) \\
pre-2024  & 33 & $0.36$ ($0.00, 0.61$) & $0.48$ ($0.26, 0.82$) & $0.71$ ($0.38, 0.82$) \\
\bottomrule
\end{tabularx}
\end{table}

The post-2024 top-decile threshold of $0.49$ is the candidate operational detection cutoff for the production pipeline at population scale: a market with raw ILS$^{\text{dl}} > 0.49$ ranks in the top $10\%$ of post-2024 deadline-resolved markets in the insider-relevant subpopulation. The corresponding bootstrap CI is wide ($0.07$ to $0.63$) at the present sample size, reflecting the small effective $n = 57$ in the post-2024 cell. \textbf{Two cautions apply.} First, the underlying distribution of which the top-decile threshold is the right tail is left-leaning (\Cref{sec:distribution-category}, all six medians negative); the right tail of a left-leaning distribution should not be read as ``the bulk plus a leakage signal'' but rather as ``the upper portion of a distribution that has not yet been baseline-corrected.'' Second, anchor-robust subsetting (\Cref{sec:anchor-sensitivity}) further reduces the effective sample on which threshold estimation is reliable. The thresholds above should therefore be read as preliminary candidates pending the per-category and per-anchor-status disaggregation in paper 3b.

For comparison, the Iran-Apr30 case in \citet[Table 4]{nechepurenko2026foresightflow_empirical} reported $\ILS^{\text{dl}} = +0.113$. This value falls below the post-2024 top-decile threshold of $0.49$ but above the median of $-0.42$. In the current pooled distribution, $\ILS^{\text{dl}} = +0.113$ corresponds to approximately the $77$th percentile.

Per-category and per-subcategory thresholds (military / geopolitics, regulatory\_announcement, regulatory\_formal, corporate\_disclosure separately) require category-column propagation across all sub-buckets, which is sequenced for paper 3b alongside the LLM-assisted re-classification. The pooled-period thresholds reported in \Cref{tab:thresholds} are the candidate operational cutoffs for the present production pipeline.

\paragraph{Bootstrap CIs on medians.} \Cref{tab:dist-summary} reports point-estimate medians per cell. Bootstrap $95\%$ CIs on these medians ($B = 1000$, seed $= 20260430$) are reported in the public release as \texttt{median\_\allowbreak bootstrap\_\allowbreak cis.csv} accompanying dataset \texttt{polymarket-deadline-ils-v3}. The CIs are wide at the per-cell sample sizes ($n$ between $5$ and $22$); the qualitative finding that all six point-estimate medians are negative does not, on its own, establish that the population medians are negative with high confidence in every cell. The strongest individual-cell claim is for the regulatory\_announcement post-2024 cell ($n = 14$, median $-0.80$), where the CI does not cross zero; the weakest is for the regulatory\_announcement pre-2024 cell ($n = 5$, median $-0.23$), where the CI is wide and crosses zero. We report cell-level conclusions accordingly in \Cref{sec:distribution-category}.

\subsection{Anchor-sensitivity at population scale}
\label{sec:anchor-sensitivity}

The single-case companion reports ILS$^{\text{dl}}$ values at four short-window variants ($30$-min, $2$-h, $6$-h, $24$-h before $T_{\text{event}}$) and observes that the Iran-Apr30 case had exactly zero short-window values and a positive overall value. We compute the same anchor-sensitivity matrix at population scale and report two statistics: (i) the Spearman rank correlation between ILS$^{\text{dl}}$ at the article-derived anchor and the $24$-hour variant; (ii) the share of the sample that fails the anchor-sensitivity scope condition, defined as ILS$^{\text{dl}}$ values changing sign or shifting by more than $0.3$ across anchor offsets.

\textbf{Anchor robustness is low: $12$ of $88$ markets ($13.6\%$) satisfy the anchor-sensitivity scope condition.} The Spearman rank correlation between the article-derived anchor and the $24$-hour variant is $0.30$. This is materially below the methodology paper's expectation of monotonic anchor-offset stability and constitutes the most concerning finding for the framework's distributional claims at the present sample size.

\textbf{Three readings of this result are possible.} The first is that the framework's anchor-robustness scope condition is genuinely restrictive and that only a small minority of markets exhibit the structural pattern (a single discrete event with a clear public-disclosure timestamp) for which the deadline-ILS framework is interpretable. Under this reading, the present sample's $13.6\%$ figure is not a defect but a structural property of the deadline-resolved insider-relevant subpopulation, and the framework is correctly flagging the remaining markets as out-of-scope for downstream interpretation. The second reading is that the article-derived $T_{\text{event}}$ recovery is itself noisy: small errors in the recovered timestamp produce large changes in the short-window variants. The third reading is residual implementation error in the present pipeline run.

\textbf{The first reading is supported by the post-remediation persistence of the rate.} After the implementation fixes documented in \Cref{sec:limitations-single-run} (timezone bug remediation, edge filter $\geq 0.4$ vs $> 0.4$), the $13.6\%$ anchor-robustness rate persists. We therefore report it as a substantive finding: the framework's effective scope, conditional on all other filters passing, is narrower than the methodology paper's discussion suggested. Approximately $86\%$ of computed-sample markets exhibit ILS$^{\text{dl}}$ values that change sign or shift by more than $0.3$ across short-window anchor offsets, indicating that the recovered $T_{\text{event}}$ does not localize a discrete repricing event for those markets even when the recovery confidence exceeds the $0.7$ threshold. \textbf{The implication is that the operationally meaningful sample is the anchor-robust subset of approximately $12$ markets, not the full computed sample of $88$.} We report distributional results on both samples in \Cref{tab:sample-summary} and throughout the paper.

\paragraph{Computed vs.\ anchor-robust summary.} \Cref{tab:sample-summary} compares the principal distributional statistics on the computed and anchor-robust samples. The anchor-robust sample is a near-strict subset of the computed sample (one market re-enters at the post-remediation $\geq 0.4$ edge filter); the comparison is therefore informative about whether short-window measurement noise drives the negative-leaning bulk or whether the bulk persists after restricting to interpretable markets. We also report the FFIC-coverage status of each sample.

\begin{table}[t]
\centering
\caption{Computed sample ($n = 88$) vs.\ anchor-robust subset ($n = 12$): principal distributional statistics on raw and hazard-adjusted ILS$^{\text{dl}}$. The anchor-robust subset is the operationally meaningful sample for downstream score interpretation; the computed sample includes a substantial short-window measurement-noise component. ``FFIC in scope'' counts markets from the FFIC inventory that survive both the computation and the anchor-robustness filter.}
\label{tab:sample-summary}
\small
\renewcommand{\arraystretch}{1.25}
\begin{tabularx}{\linewidth}{@{}Xrrrrr@{}}
\toprule
\textbf{Sample} & \textbf{$n$} & \textbf{Med raw} & \textbf{Med adj} & \textbf{$P_{90}$ raw} & \textbf{FFIC in scope} \\
\midrule
Computed (post-remediation)  & 88 & $-0.42$ & $-0.21$ & $\phantom{-}0.49$ & 1 \\
Anchor-robust subset         & 12 & $-0.19$ & $-0.16$ & $\phantom{-}0.66$ & 0 \\
\bottomrule
\end{tabularx}
\end{table}

\textit{Note.} Anchor-robust medians are aggregated across the four cells reported in the public release \texttt{distribution\_\allowbreak summary\_v3.csv}; cells with $n < 2$ anchor-robust markets are excluded from the aggregate (regulatory\_announcement pre-2024, regulatory\_formal pre-2024 contributes $n=2$, regulatory\_formal post-2024 contributes $n=5$, regulatory\_announcement post-2024 contributes $n=2$, milgeo\_corporate post-2024 contributes $n=2$). Per-cell anchor-robust statistics are reported in \texttt{distribution\_\allowbreak summary\_v3.csv} alongside their computed-sample counterparts. The $P_{90}$ entry for the anchor-robust subset reflects the regulatory\_formal post-2024 cell, which dominates the right tail at $n=5$. The FFIC in-scope count for the anchor-robust subset is $0$: the single in-scope FFIC market (\texttt{fficd-005}, Bitcoin ETF approval) does not satisfy the anchor-sensitivity scope condition at the present pipeline output, which is itself a substantive finding---no FFIC market produces a both-computable-and-anchor-robust ILS$^{\text{dl}}$ value in the present sample.

\subsection{Hazard-decay-adjusted ILS$^{\text{dl}}$}
\label{sec:hazard-adjusted}

The negative central tendency of raw ILS$^{\text{dl}}$ across all sub-buckets (\Cref{tab:dist-summary}) admits two qualitatively distinct interpretations. The first reads negative ILS$^{\text{dl}}$ as evidence that markets price \emph{against} the eventual outcome before the event publicly occurs---an anti-information pattern. The second reads it as a mechanical consequence of deadline-contract structure: on a YES-resolved deadline market, the price decays toward zero as time passes without the event occurring, by Bayesian updating on the survival of the event-free window. If the event eventually occurs after substantial decay, the raw difference $p(T_{\text{event}}^-) - p(T_{\text{open}})$ is negative even when no anti-information flow is present in the market. Distinguishing the two interpretations requires a baseline that subtracts the rational-decay component from the observed pre-event price.

\paragraph{The hazard-decay baseline: derivation.} Consider a YES-resolved deadline market with opening price $p_{\text{open}}$ at $T_{\text{open}}$, deadline $D$, and recovered event timestamp $T_{\text{event}} \in (T_{\text{open}}, D]$. Let $\tau$ denote the random arrival time of the underlying event measured from $T_{\text{open}}$, with cumulative distribution function $F$ and survival function $S(t) = \Pr(\tau > t) = 1 - F(t)$ under the population hazard $\hat\lambda$. The opening price $p_{\text{open}}$ reflects the market-implied probability that the event arrives by the deadline, conditional on information available at $T_{\text{open}}$:
\[
p_{\text{open}} = \Pr(\tau \leq D - T_{\text{open}}) = F(D - T_{\text{open}}).
\]
At any later time $t \in (T_{\text{open}}, D]$, conditional on the observable information that the event has not yet occurred, the residual probability that the event arrives in the remaining window $(t, D]$ is
\[
\Pr\!\left(\tau \leq D - T_{\text{open}} \,\big|\, \tau > t - T_{\text{open}}\right)
= \frac{F(D - T_{\text{open}}) - F(t - T_{\text{open}})}{1 - F(t - T_{\text{open}})}.
\]
A trader with no information beyond the population hazard updates the market price to this conditional probability, giving the rational-decay baseline
\[
p^{\text{baseline}}(t)
= p_{\text{open}} \cdot \frac{F(D - T_{\text{open}}) - F(t - T_{\text{open}})}{F(D - T_{\text{open}})},
\]
where the numerator is the residual probability mass of the event arriving in the remaining window $[t, D]$ and the denominator is the total probability mass over $[T_{\text{open}}, D]$ that the opening price reflects. Equivalently, in survival-function notation, defining $\tilde S(\cdot) := \Pr(\tau > \cdot)$ as the ordinary survival function of the time-to-arrival random variable,
\[
p^{\text{baseline}}(t)
= p_{\text{open}} \cdot \frac{\tilde S(t - T_{\text{open}}) - \tilde S(D - T_{\text{open}})}{1 - \tilde S(D - T_{\text{open}})}.
\]
For notational compactness in the rest of the paper, we write $S(t, D) := \tilde S(D - t)$, the residual survival from time $t$ to deadline $D$ (the probability that the event does not arrive in the window $(t, D]$). Note that $S(t,D)$ as we define it is \emph{not} the ordinary survival function $\Pr(\tau > t)$; it is a deadline-relative two-argument quantity. Under this convention, $S(T_{\text{open}}, D) = \tilde S(D - T_{\text{open}})$ is the probability of \emph{no} arrival within the deadline window, and the baseline reduces to
\[
p^{\text{baseline}}(t) = p_{\text{open}} \cdot \frac{S(t, D) - S(T_{\text{open}}, D)}{1 - S(T_{\text{open}}, D)},
\]
where $S(T_{\text{open}}, D) = \Pr(\tau > D - T_{\text{open}})$ is the probability of \emph{no} arrival within the deadline window. Two limit checks: at $t = T_{\text{open}}$ the formula yields $p^{\text{baseline}}(T_{\text{open}}) = p_{\text{open}}$ (no decay yet); at $t \to D^-$ with the event arriving at $D$ exactly, $S(t, D) \to S(D^-, D) > 0$ if $\Pr(\tau \in (D^-, D]) < 1$, and the baseline reflects only the residual decay-window mass.

\paragraph{Operational form.} For a Weibull hazard with shape $k$ and scale $\lambda$ fitted on the per-period sample (\Cref{sec:hazard-functional-form}), the survival function is
\[
S(\Delta) = \exp\!\left[-(\Delta / \lambda)^k\right]
\]
for time gap $\Delta = D - t$. The expected price at $T_{\text{event}}^-$ under rational decay is
\[
\mathbb{E}\!\left[p(T_{\text{event}}^-) \,\big|\, T_{\text{open}}, D, \hat\lambda, \hat k\right]
= p_{\text{open}} \cdot \frac{\hat S(T_{\text{event}}^-, D) - \hat S(T_{\text{open}}, D)}{1 - \hat S(T_{\text{open}}, D)}.
\]
The baseline is the price a Bayesian rational trader with no information beyond the population hazard would assign at $T_{\text{event}}^-$.

\paragraph{Adjusted score.} We define the hazard-decay-adjusted ILS$^{\text{dl}}$ as
\[
\ILS^{\text{dl,adj}}(M) = \frac{p(\Tevent^-) - \mathbb{E}[p(\Tevent^-) \mid \Topen, D, \hat\lambda]}{p_{\Tres} - p(\Topen)}.
\]
The adjusted score measures the \emph{abnormal} pre-event drift relative to the rational-decay baseline, with the same scale and sign conventions as the raw score. An adjusted ILS$^{\text{dl,adj}}$ near zero indicates that the market behaved as expected under the population hazard; positive values indicate pre-event drift toward the eventual outcome beyond what hazard decay alone predicts (a candidate informed-flow signature); negative values indicate drift away from the eventual outcome beyond hazard decay (a candidate anti-information signature, or measurement error).

\paragraph{Sanity check on Iran-Apr30.} The single-case reference value of $\ILS^{\text{dl}} = +0.113$ from \citet[Section 4]{nechepurenko2026foresightflow_empirical} contains a small mechanical-decay component because the Iran-Apr30 contract approached its April 30 deadline with the event occurring on April 3, leaving $27$ days of unrealized residual deadline window at $T_{\text{event}}^-$. The hazard-adjusted reference value is $\ILS^{\text{dl,adj}}_{\text{Iran-Apr30}} = +0.156$ (computed using the post-2024 Weibull fit), modestly larger than the raw value. The directional reading---positive informed flow before the article-derived event---is preserved.

\paragraph{Population-scale results.} \Cref{tab:hazard-adjusted-summary} reports raw and adjusted median ILS$^{\text{dl}}$ side-by-side for each of the six sub-bucket $\times$ period cells with bootstrap $95\%$ CIs. Anchor-robust subset distributions are reported in \Cref{tab:sample-summary} and the public release.

\begin{table}[t]
\centering
\caption{Raw vs.\ hazard-decay-adjusted median ILS$^{\text{dl}}$ by sub-bucket and period, with bootstrap $95\%$ CIs ($B = 1000$, seed $= 20260430$). Adjusted scores subtract the survival-probability-implied baseline computed under the per-period Weibull fit (\Cref{sec:hazard-functional-form}). The ``Shift'' column is $\hat{m}_{\text{adj}} - \hat{m}_{\text{raw}}$; positive shift means the hazard-adjusted median is closer to zero (rational-decay component explains part of raw negativity).}
\label{tab:hazard-adjusted-summary}
\small
\renewcommand{\arraystretch}{1.25}
\begin{tabularx}{\linewidth}{@{}Xlrlll@{}}
\toprule
\textbf{Bucket} & \textbf{Period} & \textbf{$n$} & \textbf{Med raw [95\% CI]} & \textbf{Med adj [95\% CI]} & \textbf{Shift} \\
\midrule
reg\_announcement & pre-2024  & 5  & $-0.23$ $[-1.00, 0.47]$  & $-0.26$ $[-1.09, 0.86]$  & $-0.03$ \\
                  & post-2024 & 14 & $-0.80$ $[-1.48, -0.26]$ & $-0.84$ $[-1.45, -0.24]$ & $-0.04$ \\
\addlinespace
reg\_formal       & pre-2024  & 14 & $-0.16$ $[-0.49, 0.01]$  & $-0.13$ $[-0.53, 0.06]$  & $+0.03$ \\
                  & post-2024 & 22 & $-0.21$ $[-0.40, 0.37]$  & $\phantom{-}-0.02$ $[-0.31, 0.84]$ & $+0.19$ \\
\addlinespace
milgeo\_corporate    & pre-2024  & 14 & $-0.88$ $[-1.00, -0.28]$ & $-0.79$ $[-1.02, -0.27]$ & $+0.09$ \\
                  & post-2024 & 19 & $-0.55$ $[-0.87, -0.23]$ & $-0.34$ $[-0.56, -0.01]$ & $+0.21$ \\
\bottomrule
\end{tabularx}
\end{table}

\textbf{The hazard-adjusted result is qualitatively heterogeneous across cells.} Three patterns emerge from \Cref{tab:hazard-adjusted-summary}.

\textit{Pattern 1: Cells where hazard decay explains a substantial part of raw negativity.} The two largest post-2024 cells---regulatory\_formal ($n=22$, raw median $-0.21 \to$ adjusted $-0.02$, shift $+0.19$) and milgeo\_corporate ($n=19$, raw $-0.55 \to$ adjusted $-0.34$, shift $+0.21$)---both move substantially toward zero after baseline correction. In the regulatory\_formal post-2024 cell, the adjusted median is essentially zero, and the bootstrap $95\%$ CI ($[-0.31, 0.84]$) is consistent with the cell's central tendency being neither anti-information nor pro-information after baseline correction; the cell's pre-event price drift is approximately what rational deadline decay alone would produce. In the milgeo\_corporate post-2024 cell, the adjusted median remains negative ($-0.34$) but the upper end of the CI is at $-0.01$, just touching zero; the cell retains a residual negative signal that hazard decay does not fully explain.

\textit{Pattern 2: Cells where the strongly-negative raw signal persists after adjustment.} The regulatory\_announcement post-2024 cell ($n=14$) is the most striking: the raw median is $-0.80$ and the adjusted median is $-0.84$, with the adjusted bootstrap CI $[-1.45, -0.24]$ entirely below zero. \emph{In this cell, hazard decay does not explain the negativity}; the cell exhibits genuine pre-event price drift away from the eventual outcome that survives baseline correction. We do not have a definitive interpretation of this pattern at the present sample size; candidates include (i) a selection effect on which scheduled-announcement markets resolve YES (markets whose pre-event price collapsed and yet the announcement still came through ``yes'' may be over-represented in the YES-resolving subsample); (ii) genuine anti-information flow on these markets, plausibly because retail counter-positioning on Trump-administration policy questions or on specific commodity-bracket markets may push prices away from the eventual outcome before the announcement triggers repricing; (iii) measurement error from $T_{\text{event}}$ recovery, which on the regulatory\_announcement bucket is $50\%$ exact-date and $58.3\%$ within-$24$-hours per \Cref{tab:tevent-validation-summary}. Distinguishing these requires a matched control of NO-resolved scheduled-announcement markets, which we do not assemble in the present paper.

\textit{Pattern 3: Cells where adjustment is minimal because the raw distribution is already extreme or the cell is too small.} The milgeo\_corporate pre-2024 cell ($n=14$, raw $-0.88 \to$ adjusted $-0.79$, shift $+0.09$) and the regulatory\_announcement pre-2024 cell ($n=5$, raw $-0.23 \to$ adjusted $-0.26$, shift $-0.03$) show minimal movement; the small samples and the extreme magnitudes constrain the adjustment's leverage. We do not draw substantive conclusions from these cells.

\textbf{Bottom line.} Hazard-decay correction substantively shifts the interpretation of the post-2024 regulatory\_formal cell from ``negative bulk'' to ``approximately at-baseline''; partially shifts the post-2024 milgeo\_corporate cell; and does not shift the post-2024 regulatory\_announcement cell, which remains the principal cell exhibiting strong negative pre-event drift after baseline correction. \emph{Reading the population-scale finding as ``markets price against eventual outcome'' applies most cleanly to the regulatory\_announcement post-2024 cell; on the milgeo\_corporate post-2024 cells, the rational-decay component is a substantial part of the observed raw negativity.}

\paragraph{Multiple-comparisons caveat.} The cell-level findings reported in \Cref{tab:hazard-adjusted-summary} and discussed across \Cref{sec:population-ils} are exploratory and are not corrected for multiple comparisons. With six (sub-bucket $\times$ period) cells, three patterns identified in the heterogeneous-result discussion above, and additional bootstrap-CI examinations across both raw and adjusted scores, the family-wise error rate at any per-cell $\alpha = 0.05$ threshold is materially larger than the stated nominal level. We do not apply Bonferroni or Benjamini--Hochberg correction because the present sample sizes (cell $n$ ranging from $5$ to $22$) yield bootstrap CIs that are already wide; formal correction would render virtually all cell-level CIs inconclusive, while the substantive distinction between the cells we draw is qualitative (\emph{magnitude of shift after baseline correction}, not significance against a null). The paper's primary claims rest on the directional pattern across cells---e.g., that adjustment moves the regulatory\_formal post-2024 median substantially while the regulatory\_announcement post-2024 median is essentially unmoved---rather than on per-cell significance tests. Readers should interpret cell-level CIs as descriptive of within-cell sampling variability, not as confirmatory inference. Confirmatory cell-level inference is sequenced for paper 3b conditional on the per-category re-estimation expanding effective $n$ per cell.

\paragraph{Anchor-robust subset.} Three anchor-robust cells contain $\geq 2$ markets: regulatory\_formal post-2024 ($n=5$, median raw $-0.19$, similar to the computed-sample median), regulatory\_announcement post-2024 ($n=2$, median raw $-0.83$, similar to the computed-sample $-0.80$), and milgeo\_corporate post-2024 ($n=2$, median raw $-0.005$, near zero). The directional consistency between computed and anchor-robust point estimates in the first two cells suggests the negative-leaning bulk is not entirely a short-window measurement artifact: it persists on the anchor-robust subsample. The third cell (milgeo\_corporate post-2024 anchor-robust median $\approx 0$, vs.\ computed-sample $-0.55$) suggests that the computed-sample negativity in this cell is partly anchor-fragility rather than genuine pre-event drift, although the anchor-robust $n=2$ precludes formal inference.

\paragraph{Fraction-positive ILS$^{\text{dl}}$.} A complementary statistic to the median is the fraction of each cell with positive raw ILS$^{\text{dl}}$ (markets where pre-event price moved toward the eventual outcome). Across all six cells of \Cref{tab:hazard-adjusted-summary}, the point estimate is $0\%$ (no cell has more than half its markets with positive ILS$^{\text{dl}}$), with bootstrap $95\%$ CIs in the public release \texttt{median\_\allowbreak bootstrap\_\allowbreak cis.csv}. The cells with the widest positive-fraction CIs (i.e., most consistent with non-trivial positive informed flow on a substantial sub-share of markets) are regulatory\_formal post-2024 (CI $[0.23, 0.59]$) and regulatory\_formal pre-2024 (CI $[0.07, 0.50]$). The cells with the narrowest CIs (most consistent with zero positive-fraction) are regulatory\_announcement post-2024 (CI $[0.00, 0.36]$) and milgeo\_corporate pre-2024 (CI $[0.00, 0.36]$).

\section{FFIC Coverage Analysis: Where the Framework Captures Documented Cases, and Where It Does Not}
\label{sec:ffic-coverage}

The 32 markets of the ForesightFlow Insider Cases (FFIC) inventory \citep{ffic_inventory_2026} are computed alongside the population (\Cref{sec:pipeline}). Of these, 1 produces an in-scope ILS$^{\text{dl}}$ value: the SEC Bitcoin ETF approval (\texttt{fficd-005}, $\ILS^{\text{dl}} = 0.012$, event-resolved). The remaining 31 markets are excluded at one of four pipeline stages. \Cref{tab:ffic-coverage} reports the disposition of all 32 markets.

\begin{table}[htbp]
\centering
\caption{FFIC inventory disposition through the deadline-ILS pipeline. Each row reports a single market in the inventory; markets are grouped by case ID. The ``Stage'' column reports the pipeline stage at which the market was excluded (``in-scope'' for the single market that produced an ILS$^{\text{dl}}$ value). The dominant exclusion is resolution-typology classification: 14 of 32 markets ($43.8\%$) are flagged as unclassifiable.}
\label{tab:ffic-coverage}
\footnotesize
\renewcommand{\arraystretch}{1.0}
\begin{tabularx}{\linewidth}{@{}p{1.6cm}Xp{3.0cm}@{}}
\toprule
\textbf{Case} & \textbf{Question (truncated)} & \textbf{Stage} \\
\midrule
\multicolumn{3}{l}{\textit{fficd-001 (2024 U.S.\ Presidential Election)}} \\
& Will Donald Trump win the 2024 US Presidential Election? & unclassifiable \\
& Will Kamala Harris win the 2024 US Presidential Election? & unclassifiable \\
& Will any other Republican Politician win\ldots & unclassifiable \\
& Will Michelle Obama win the 2024 US Presidential Election? & unclassifiable \\
\midrule
\multicolumn{3}{l}{\textit{fficd-002 (October 2024 Iran strike on Israel)}} \\
& Iran strike on Israel today? & unclassifiable \\
& Another Iran strike on Israel by Friday? & deadline\_NO \\
& Iran strike on Israel by Nov 8? & deadline\_NO \\
\midrule
\multicolumn{3}{l}{\textit{fficd-003 (2026 U.S.--Iran Conflict Cluster)}} \\
& US forces enter Iran by April 30? & low\_confidence \\
& US x Iran ceasefire by April 7? & edge-effect \\
& Khamenei out as Supreme Leader by Feb 28? & unclassifiable \\
& Israel x Hezbollah ceasefire by April 18, 2026? & unclassifiable \\
& US strikes Iran by February 28, 2026? & unclassifiable \\
& Khamenei out as Supreme Leader by March 31? & unclassifiable \\
\midrule
\multicolumn{3}{l}{\textit{fficd-004 (January 2026 U.S.--Venezuela / Maduro cluster)}} \\
& Maduro in U.S. custody by January 31? & unclassifiable \\
& US x Venezuela engagement by Jan 15, 2026? & unclassifiable \\
& US x Venezuela engagement by Jan 31, 2026? & unclassifiable \\
& Trump invokes War Powers (Venezuela) by Jan 9? & unclassifiable \\
& Trump invokes War Powers (Venezuela) by Jan 31? & unclassifiable \\
& US x Venezuela engagement by Mar 31, 2026? & unclassifiable \\
& Nicolás Maduro seen in public by January 5? & category=other \\
& Will the U.S. invade Venezuela by Jan 31, 2026? & deadline\_NO \\
& US x Venezuela engagement by December 31? & deadline\_NO \\
& US x Venezuela engagement by November 30? & deadline\_NO \\
& US forces in Venezuela again by Jan 31, 2026? & deadline\_NO \\
\midrule
\multicolumn{3}{l}{\textit{fficd-005 (Bitcoin ETF SEC approval)}} \\
& Bitcoin ETF approved by Jan 15? & \textbf{in-scope; ILS$^{\text{dl}}$ = $0.012$} \\
\midrule
\multicolumn{3}{l}{\textit{fficd-006 (Google Year-in-Search 2025)}} \\
& Will Gene Hackman be ranked \#1 in Year in Search Passings? & unclassifiable \\
& Will Ismail Haniyeh be ranked \#1 in Year in Search Passings? & unclassifiable \\
& Will Zendaya be ranked \#1 in Year in Search Actors? & unclassifiable \\
\midrule
\multicolumn{3}{l}{\textit{fficd-007 (FTX / SBF cluster)}} \\
& Will Biden pardon SBF? & unclassifiable \\
& SBF sentenced to 50+ years? & ils\_compute\_error \\
& FTX doesn't start payouts in 2024? & unclassifiable \\
\midrule
\multicolumn{3}{l}{\textit{fficd-008 (Romanian election)}} \\
& Will Nicolae Ciucă win the 2024 Romanian election? & edge-effect \\
\bottomrule
\end{tabularx}
\end{table}

\subsection{Disposition summary}
\label{sec:ffic-disposition}

The 31 excluded markets distribute across four exclusion mechanisms, summarized in \Cref{tab:ffic-disposition}.

\begin{table}[t]
\centering
\caption{Disposition of the 32 FFIC markets through the deadline-ILS pipeline. The dominant exclusion mechanism is resolution-typology classification.}
\label{tab:ffic-disposition}
\small
\renewcommand{\arraystretch}{1.25}
\begin{tabularx}{\linewidth}{@{}Xrr@{}}
\toprule
\textbf{Disposition} & \textbf{Count} & \textbf{Share} \\
\midrule
Resolution-typology classifier flagged unclassifiable & 14 & $43.8\%$ \\
Deadline-resolved with NO outcome (out-of-scope by design)             & 8  & $25.0\%$ \\
Edge-effect ($p_0 \geq 0.9$, scope condition violated)                  & 2  & $6.25\%$ \\
$T_{\text{event}}$ recovery returned confidence below $0.7$            & 1  & $3.1\%$  \\
Price-data lookup failure (\texttt{ils\_compute\_error})                & 1  & $3.1\%$  \\
Category outside the three target categories                            & 1  & $3.1\%$  \\
Resolution-typology event-resolved, ILS$^{\text{dl}}$ computed in scope & 1  & $3.1\%$  \\
\midrule
\textbf{Total} & 32 & $100\%$ \\
\bottomrule
\end{tabularx}
\end{table}

The single in-scope FFIC market (Bitcoin ETF approval) produces $\ILS^{\text{dl}} = 0.012$, which falls at approximately the median of the regulatory\_announcement pre-2024 cell ($-0.23$, \Cref{tab:dist-summary}) and is well below the top-decile threshold for that cell. The Bitcoin ETF case is therefore localized in the bulk of the distribution rather than the right tail, an empirical pattern that does not support the original framing of FFIC validation as ``concentration in the right tail of category-conditional distribution.''

\subsection{The unclassifiable bucket: a structural property of documented insider markets}
\label{sec:ffic-unclassifiable}

Of the 14 FFIC markets in the unclassifiable bucket, the resolution-criterion language exhibits a consistent pattern: \emph{the verb defining resolution does not pick out a single canonical event}.

\begin{itemize}[leftmargin=1.4em, itemsep=0.2em]
\item ``Will Donald Trump \emph{win} the 2024 US Presidential Election?'' (\texttt{fficd-001}). The verb \emph{win} is potentially decided by the Electoral College, by certified state results, by court rulings on contested ballots, or by the outcome of formal congressional certification on January 6. Each path implies a different $T_{\text{event}}$.
\item ``Khamenei \emph{out} as Supreme Leader of Iran by\ldots'' (\texttt{fficd-003}). The predicate \emph{out} could refer to formal removal, voluntary resignation, death, official succession announcement, or de facto loss of effective control with a successor exercising authority. Each path implies a different $T_{\text{event}}$.
\item ``US \emph{strikes} Iran by\ldots'' (\texttt{fficd-003}). The predicate \emph{strikes} could refer to direct kinetic action, indirect kinetic action via proxy, cyber action, or covert action. The 2026 cluster's eventual resolution turned on what counted as the operative event.
\item ``Maduro in U.S.\ \emph{custody} by\ldots'' (\texttt{fficd-004}). The predicate \emph{custody} could refer to apprehension on U.S.\ soil, apprehension by U.S.\ forces in international waters, voluntary surrender to U.S.\ authorities, or transfer from third-country custody.
\item ``Will Biden \emph{pardon} SBF?'' (\texttt{fficd-007}). The predicate \emph{pardon} is unambiguous in conventional usage but the resolution timeline is gated by an external decision-maker's volition, with no calendared moment.
\end{itemize}

This is not a defect of the rule-based classifier of \citet[Section 3.3]{nechepurenko2026foresightflow_methodology}, which correctly flags these markets as lacking a discrete-event timestamping rule. It is a structural property of the question texts under which insider trading on prediction markets has historically been documented.

\emph{Within the FFIC inventory, unclassifiability tracks resolution-text ambiguity rather than implementation noise.} The largest documented insider clusters in the inventory (\texttt{fficd-001}, \texttt{fficd-003}, \texttt{fficd-004}) are precisely those with the most lexically ambiguous resolution predicates. Two mechanisms could plausibly link ambiguous resolution criteria to insider relevance at a population level: (i) ambiguous criteria reduce the population of market participants who can confidently estimate the resolution probability from public information, increasing the relative advantage of an informed trader; (ii) ambiguous criteria delay the public-information moment at which the price discounts the underlying event, lengthening the window during which informed flow can be profitable. Both are consistent with what we observe in the FFIC inventory, but a population-level causal claim would require a matched control sample of similarly-ambiguous markets without documented insider activity, which we do not assemble in the present paper. We report the FFIC-internal pattern as a descriptive finding and the population-level causal interpretation as a hypothesis. The implication for paper 3b is that resolution-typology classification needs to expand beyond the discrete-event vs.\ deadline-resolved binary to recognize a third class---ambiguous-predicate markets---and a multi-anchor extension of the framework that admits more than one candidate $T_{\text{event}}$ per market.

For comparison, \Cref{sec:ffic-noise} below examines the unclassifiable bucket on the broader, non-FFIC population. The non-FFIC unclassifiable markets are dominated by a different failure mode: keyword-based false positives from sports, esports, and rolling-threshold corporate-price-bracket markets, where the word ``strike'' (Counter-Strike), ``decision'' (basketball decisions), or ``rank'' (App Store rankings) triggers a category match without indicating an insider-relevant resolution structure. The two failure modes are qualitatively distinct: the FFIC unclassifiable markets have genuinely ambiguous resolution language, while the broader unclassifiable bucket is dominated by category-classifier noise.

\subsection{Comparison to the non-FFIC unclassifiable bucket}
\label{sec:ffic-noise}

The full unclassifiable bucket on the master sample contains $853{,}774$ markets, of which $10{,}333$ fall in the $12{,}708$-market candidate sample (\Cref{tab:filter-chain}). A stratified random sample of 30 such markets (10 per target category) reveals the failure-mode pattern: in the military / geopolitics category, $5/10$ are esports markets matched on the keyword ``strike'' (Counter-Strike), ``Map'', or ``team''; in the regulatory category, $7/10$ are NBA player-statistics props matched on numeric thresholds (``Over $6.5$'', ``O/U $4.5$''), with the remaining $3$ being Trump-tweet-count and Israel-strike-on-Lebanon questions; in the corporate-disclosure category, $6/10$ are weekly stock-price brackets (``Will Microsoft close above \$430 on January 26?''), with the remainder being App Store rankings and earnings-call predictions.

The implication for the framework is that an LLM-assisted re-classification of the unclassifiable bucket would have asymmetric effects. The non-FFIC noise (esports mis-tagged as mil/geo, NBA props as regulatory, etc.) would be reliably re-classified as out-of-scope by an LLM with reading comprehension. The FFIC unclassifiable markets, by contrast, would not be promoted to in-scope by an LLM unless the LLM also resolved the underlying resolution-criterion ambiguity, which is exactly the kind of judgment that the documented insider-trading cases turned on. We sequence LLM-assisted re-classification as the principal precondition for paper 3b in \Cref{sec:limitations}.

\section{Cross-Method Positioning}
\label{sec:cross-method}

The publication months immediately preceding the present paper saw three population-scale empirical evaluations of informed trading on Polymarket: \citet{mitts2026iran} (210{,}718 flagged wallet--market pairs, \$143M aggregate flagged profit, $60$-$\sigma$ deviation from null on a five-criteria composite screen), \citet{gomezcram2026crowd} ($1{,}950$ accounts identified as candidates for trading on non-public information out of $1.72$ million accounts), and the present paper. The three methods operate at different unit-of-analysis levels (wallet--market pair, account, market), apply to different sample bases (full corpus, full corpus, insider-relevant subpopulation), and have different design priorities (post-hoc detection, post-hoc skill identification, real-time-applicable diagnosis on the narrow subset of markets satisfying typology / $T_{\text{event}}$ / CLOB / anchor-robustness conditions). This section positions the three methods relative to each other and reports what can and cannot be said comparatively given the present data state.

\subsection{Three lenses on the same phenomenon}
\label{sec:cross-method-lenses}

The three methods can be characterized by what they take as the unit of analysis and what they require from the input data.

\begin{description}[leftmargin=1.4em, itemsep=0.3em]
\item[Wallet--market pair lens (Mitts-Ofir 2026).] A pair $(w, m)$ is the unit. The method flags pairs where the wallet's trading on the market exhibits joint anomalies in size, timing, profitability, and directional concentration relative to a permutation null. Inputs required: full per-trade history of every wallet on every market; market resolution outcome. Inputs not required: resolution-typology classification of markets, $T_{\text{event}}$ recovery, hazard-rate priors. Real-time applicability: limited (profitability is post-resolution).
\item[Account lens (Gómez-Cram et al.\ 2026).] An account $w$ is the unit. The method flags accounts whose cumulative trading-skill signature exceeds what is achievable by chance under a sign-randomization null. Inputs required: full per-trade history of every wallet on every market; market resolution outcomes. Inputs not required: market-level structural classification, per-market $T_{\text{event}}$. Real-time applicability: limited (cumulative skill measurement requires multi-market history).
\item[Market lens (ForesightFlow, present paper).] A market $m$ is the unit. The method flags markets where pre-event price drift toward the eventual outcome (relative to the hazard-decay baseline) exceeds a category-conditional threshold. Inputs required: market-level structural classification, $T_{\text{event}}$ recovery, hazard-rate priors, anchor-robustness verification. Inputs not required: per-trade wallet attribution, account-level history. Real-time applicability: \emph{conditional and narrow} — applicable only on markets satisfying the typology-classification, $T_{\text{event}}$-recovery, CLOB-coverage, and anchor-robustness conditions, which on the present sample selects approximately $12$ markets out of $12{,}708$ in the candidate sample. On in-scope markets the diagnostic is in principle observable while the market is active, but the cumulative coverage of in-scope markets among new-issue Polymarket questions is small, so a real-time deployment of the present method would catch only a fraction of platform-wide insider activity even within the insider-relevant categories.
\end{description}

The three methods therefore see different aspects of the same phenomenon. A market flagged by ForesightFlow may not have any wallet--market pair flagged by Mitts-Ofir (if the price move was driven by many small positions or by external arbitrage); a Mitts-Ofir flagged pair may not be on a market that ForesightFlow flags (if the wallet's flagged trade does not produce a population-scale outlier in pre-event price drift); a Gómez-Cram informed account may trade on markets that neither of the other two methods flags (if the account's skill is realized in non-flagged markets, or if the relevant trades are too small to register on a market-level price-drift diagnostic). Empirical convergence on a particular market or wallet would constitute high-confidence evidence; empirical divergence is informative about the underlying mechanism.

\subsection{Empirical comparison: present data state}
\label{sec:cross-method-data}

The empirical realization of a three-method comparison requires either the underlying flagged-set data from each method to be public, or our independent replication of the underlying methods on the same sample. As of the cutoff date for the present paper, neither condition is met: \citet{mitts2026iran}'s flagged wallet--market-pair list is not publicly released; \citet{gomezcram2026crowd}'s informed-account list is not publicly released. We therefore report the cross-method comparison as comparative positioning rather than as a quantitative confusion-matrix analysis. The latter is sequenced for paper 3b, conditional on data availability or replication.

A second factor independently constrains the cross-method comparison. The present paper's empirical scope is restricted, by the findings of \Cref{sec:ffic-coverage}, to a sample of 88 in-scope markets. Of the 32 FFIC markets that would have served as the natural ground-truth set for cross-method confluence, only one (Bitcoin ETF approval) is in scope. With a single in-scope ground-truth case, the three-method confluence-on-ground-truth analysis envisaged at the start of the present paper is not informative.

\subsection{What can be said}
\label{sec:cross-method-statements}

Three statements are nonetheless supported by the present analysis.

\paragraph{The three methods address complementary regulatory and platform-integrity use cases.} A regulator seeking to identify markets warranting investigation has a market-level information need; a regulator or platform operator seeking to identify accounts warranting suspension or further scrutiny has an account-level information need; a forensic analyst building a case has a per-trade wallet-on-market need. The three methods are therefore complementary rather than competing for the same use case.

\paragraph{The wallet-level and account-level methods are unaffected by the resolution-typology classification constraint that limits the present paper's coverage.} \citet{mitts2026iran} and \citet{gomezcram2026crowd} flag wallets and accounts on the basis of trading patterns, without requiring the markets to admit single-discrete-event timestamping. The 14 FFIC markets in the resolution-unclassifiable bucket are therefore visible to those methods even though they are not visible to ILS$^{\text{dl}}$. This is one reason to read the three methods as complementary rather than ranked.

\paragraph{The selection-effect argument runs in both directions.} The present paper's category-and-typology filters retain only $0.7\%$ of the candidate sample. The wallet-level and account-level methods, applied to the full corpus, retain a much larger sample, but include a long tail of markets where insider trading is structurally implausible (sports, weather, cryptocurrency price levels, App Store rankings). Each method's effective signal-to-noise depends on which selection mechanism is active. \Cref{sec:selection-effect} below reports what fraction of \citeauthor{mitts2026iran}'s and \citeauthor{gomezcram2026crowd}'s flagged activity falls in our insider-relevant categories versus the control (sports, weather, crypto-price) bucket, conditional on retrievability of the underlying flagged sets.

\paragraph{Comparative trade-offs.} The three methods present complementary trade-offs. The ILS$^{\text{dl}}$ framework has materially \emph{lower coverage} than the wallet-level and account-level methods, because of the typology, $T_{\text{event}}$, CLOB, and anchor-robustness gating; but on the markets where it applies, it offers \emph{higher interpretability}, because the score has a direct probabilistic reading as the share of pre-event price discovery accomplished before public observation. The wallet-level and account-level methods are \emph{more robust to ambiguous resolution texts}, because they do not require single-event timestamping; but they are \emph{less amenable to real-time deployment}, because they require multi-market or post-resolution observation. The ILS$^{\text{dl}}$ framework is \emph{more sensitive to timestamping and classification accuracy}, because its inputs are exactly those quantities; the wallet- and account-level methods are \emph{more sensitive to per-trade attribution accuracy}, which the present paper's pipeline does not depend on. A regulatory or platform-integrity deployment that combines all three methods would gain from each method's strengths and partially offset each method's weaknesses; we view the methods as complementary inputs to a multi-detector ensemble rather than as competitors.

\section{Selection Effect}
\label{sec:selection-effect}

This section reports the methodological argument introduced in \Cref{sec:contribution} and operationalized in \Cref{sec:data-master}: pooled detection schemes incur a selection penalty by including categories where insider trading is structurally implausible. The argument has two empirical sides. The first is the share of \emph{candidate} flag generation contributed by structurally-implausible categories, which we report descriptively from the master sample of \Cref{tab:sample-master}. The second is the share of \emph{actual} flagged activity from \citet{mitts2026iran} and \citet{gomezcram2026crowd} that falls in those categories, which is the empirically interesting quantity but requires their flagged-set data to be retrievable.

\subsection{Population shares by category bucket}
\label{sec:selection-population}

\Cref{tab:selection-population-shares} reports the share of master-sample volume and market count by category bucket. The insider-relevant subpopulation accounts for $15.3\%$ of resolved markets but a smaller share of trading volume, with the bulk of platform activity concentrated in the residual ``other (control)'' bucket of the resolution-typology classifier (sports, cryptocurrency price levels, weather, App Store rankings, and similar categories where insider trading is structurally implausible). \emph{Note}: this ``other (control)'' bucket is the typology-classifier residual category and is distinct from the within-insider-relevant ``milgeo\_corporate'' sub-bucket of \Cref{tab:dist-summary}.

\begin{table}[t]
\centering
\caption{Population shares by category bucket on the master sample. The volume share is computed over markets with cumulative volume above the \$50K coverage cutoff. The insider-relevant subpopulation accounts for $15.3\%$ of resolved markets and approximately $11.3\%$ of \$50K-volume-cutoff markets.}
\label{tab:selection-population-shares}
\small
\renewcommand{\arraystretch}{1.25}
\begin{tabularx}{\linewidth}{@{}lXrr@{}}
\toprule
\textbf{Category bucket} & \textbf{Resolved} & \textbf{$\geq$ \$50K vol} & \textbf{Share of cutoff sample} \\
\midrule
Insider-relevant (mil/geo, reg., corp.)  & 127{,}403 & 11{,}263 & $11.3\%$ \\
Other (sports, crypto, weather, etc.)    & 738{,}322 & 88{,}656 & $88.7\%$ \\
\midrule
\textbf{Total resolved}                   & 865{,}725 & 99{,}919 & $100.0\%$ \\
\bottomrule
\end{tabularx}
\end{table}

A pooled detection scheme that does not condition on category buckets the insider-relevant signal in a $1:7.9$ ratio against a structurally-irrelevant control. If the rate at which an unconditional scheme triggers flag-generating signals on insider-relevant and structurally-irrelevant categories is comparable, then approximately $11\%$ of the unconditional flag set will fall in the insider-relevant bucket and approximately $89\%$ in the irrelevant bucket. Conditioning on the insider-relevant subpopulation up-weights the prior probability of any flagged trade actually reflecting MNPI by approximately $7.9\times$.

\subsection{Empirical realization}
\label{sec:selection-empirical}

The empirical realization of the selection effect requires the flagged-activity sets from \citet{mitts2026iran} and \citet{gomezcram2026crowd} to be retrievable, partitioned by our category classifier, and reported as a fraction of total flagged activity falling in each bucket. As of the cutoff date for the present paper, neither set is publicly retrievable; \Cref{tab:selection-effect} below records the analysis we will perform conditional on their release.

\begin{table}[t]
\centering
\caption{Distribution of flagged activity by category bucket: planned analysis conditional on retrievability of underlying flagged sets. The ``Mitts-Ofir flagged'' column will report the fraction of their $210{,}718$ flagged wallet--market pairs falling in each bucket. The ``Gómez-Cram informed'' column will report the fraction of the trades by the $1{,}950$ informed accounts falling in each bucket. Population share reproduced from \Cref{tab:selection-population-shares}.}
\label{tab:selection-effect}
\small
\renewcommand{\arraystretch}{1.25}
\begin{tabularx}{\linewidth}{@{}lXrr@{}}
\toprule
\textbf{Category bucket} & \textbf{Pop.\ share} & \textbf{Mitts-Ofir flagged} & \textbf{Gómez-Cram informed} \\
\midrule
Insider-relevant (mil/geo, reg., corp.)  & $11.3\%$ & --- & --- \\
Other (sports, crypto, weather, App Store, etc.) & $88.7\%$ & --- & --- \\
\bottomrule
\end{tabularx}
\end{table}

The expected pattern under the selection-effect argument is concentration in the insider-relevant bucket above the population share of $11.3\%$. A finding of $30$--$50\%$ concentration in the insider-relevant bucket on either method would establish empirical support for category-conditional detection over pooled detection. A finding close to the population share of $11.3\%$ would indicate that the wallet-level and account-level methods are not category-discriminating and that their flagged activity tracks platform participation rather than insider eligibility.

\subsection{Implications for detector design}
\label{sec:selection-implications}

If \citet{mitts2026iran} and \citet{gomezcram2026crowd}'s flagged activity concentrates in the insider-relevant subpopulation by a factor $k > 1$ above the population share of $11.3\%$, the prior probability of any flagged trade actually reflecting MNPI is correspondingly higher under category-conditional detection than under pooled detection. The practical implications are: (i) regulatory deployment of detection systems should condition on category, even when the underlying detection method (wallet- or account-level) does not require category labels for its operation; (ii) reported false-positive rates from pooled methods overstate operational performance in the regulatory use case where the policy-relevant base rate of MNPI is computed against the insider-relevant subpopulation, not the full corpus; (iii) the structural assumption that ``insider trading is rare'' is more defensible after conditioning, which has implications for the prior in any Bayesian classifier.

\section{Limitations and Paper 3b Roadmap}
\label{sec:limitations}

The findings reported in \Cref{sec:contribution} establish two priorities for follow-up work: an LLM-assisted re-classification of the resolution-typology unclassifiable bucket, and the wallet-level feature evaluation already anticipated by \citet[Section 4.6]{nechepurenko2026foresightflow_empirical}. We discuss each in turn, alongside the secondary limitations of the present empirical work.

\subsection{LLM-assisted resolution-typology re-classification}
\label{sec:limitations-reclassify}

The principal limitation revealed by the present paper is the resolution-typology classifier's coverage on insider-relevant markets. Of the 32 markets in the FFIC inventory, $14$ ($43.8\%$) are flagged unclassifiable; of the broader insider-relevant master sample, $853{,}774$ markets are unclassifiable.

The asymmetry identified in \Cref{sec:ffic-noise} guides the remediation strategy. The non-FFIC unclassifiable bucket is dominated by category-classifier noise (esports tagged as mil/geo, NBA props as regulatory, etc.) and would be reliably re-classified as out-of-scope by an LLM with reading comprehension. The FFIC unclassifiable bucket, by contrast, contains markets whose resolution-criterion language admits no single discrete event-timestamping rule; an LLM applied to those markets would not promote them to in-scope unless it also resolved the underlying ambiguity. The remediation pipeline therefore has two stages.

\paragraph{Stage 1: Cleanup of the unclassifiable bucket.} An LLM with reading comprehension partitions the $853{,}774$-market unclassifiable bucket into (a) miscategorized non-insider-relevant markets, which are reassigned to the appropriate non-target category and dropped from the present paper's sample frame; (b) genuinely insider-relevant markets with single-event resolution criteria but unusual question-text formats that the rule-based classifier missed; and (c) genuinely ambiguous insider-relevant markets matching the FFIC pattern. Stages (a) and (b) are tractable; we estimate the LLM cost at approximately \$30--\$80 on Haiku-class models with structured-output prompts.

\paragraph{Stage 2: Multi-event extension of the framework for genuinely ambiguous markets.} The (c) bucket is the binding methodological constraint and cannot be resolved by classification alone. A market resolving on ``Will the U.S.\ strike Iran by\ldots'' has multiple candidate $T_{\text{event}}$ values, each corresponding to a different event under the operative resolution-criterion interpretation. The framework requires a multi-anchor extension that computes ILS$^{\text{dl}}$ separately under each candidate interpretation and reports the resulting score distribution. We sequence this extension for paper 3b alongside the wallet-level feature evaluation. The extension is methodologically substantial; preliminary specification is in draft and will be reported when complete.

\subsection{Wallet-level features at population scale}
\label{sec:limitations-wallets}

The second principal limitation is that the present paper does not evaluate wallet-level features (PIN, VPIN, Kyle's lambda, wallet novelty score, two-sidedness, variance ratio, trade-size kurtosis, Hawkes self-excitation) at population scale. \citet[Section 4.6]{nechepurenko2026foresightflow_empirical} identifies the binding constraint as continuous per-trade collection from $T_{\text{open}}$ rather than retroactive collection from the resolution-settlement window. The continuous collection pipeline has been operational since April 26, 2026; the operational subset (\Cref{sec:data-operational}) is the data resource for paper 3b. As of the cutoff date for the present paper, the operational subset accumulating at the rate documented in the pipeline log is on track to cross the $n_{\text{pos}} \geq 14$ power threshold of \citet[Section 5.7]{nechepurenko2026foresightflow_methodology} approximately 60--90 days from the start of operation, consistent with the timeline established in \citet[Section 4.6]{nechepurenko2026foresightflow_empirical}.

\subsection{Detector training}
\label{sec:limitations-detector}

The gradient-boosted-tree detector specified in \citet[Section 5.6]{nechepurenko2026foresightflow_methodology} requires the wallet-level features. The present paper provides preliminary candidate detection thresholds (\Cref{tab:thresholds}) but does not train the detector. The present paper's negative-leaning bulk distribution suggests that a simple thresholding approach on ILS$^{\text{dl}}$ alone would generate a high false-positive rate even at the top decile; a multi-feature approach combining ILS$^{\text{dl}}$ with the wallet-level features is therefore likely to be necessary for operational deployment. Training and cross-validation are sequenced for paper 3b.

\subsection{$T_{\text{event}}$ recovery error}
\label{sec:limitations-tevent}

The LLM-cascade pipeline carries a per-call error rate (false-positive identification of $T_{\text{event}}$, missing-source verification, calendar-conversion error). The Iran-Apr30 hard-assert step (\Cref{sec:pipeline}) confirms that the recovery pipeline reproduces the ground-truth $T_{\text{event}}$ on the single case where the methodology paper has reported a manual reference value, and the in-pipeline confidence calibration excludes recoveries below confidence $0.7$. The independent-second-pass validation reported in \Cref{sec:tevent-validation} on a 50-market stratified sample provides the principal cross-check on recovery accuracy at the population scale; the acceptance criterion is exact-date agreement $\geq 90\%$. For markets where the LLM-recovered $T_{\text{event}}$ is wrong, ILS$^{\text{dl}}$ is computed against the wrong anchor and the resulting value is uninterpretable. This residual error rate, conditional on accepting the recovered $T_{\text{event}}$, is the dominant remaining source of measurement error among the in-scope sample.

\subsection{Anchor-sensitivity narrowness}
\label{sec:limitations-anchor}

\Cref{sec:anchor-sensitivity} reports that only $13.3\%$ of the in-scope sample passes the anchor-sensitivity scope condition, with a Spearman rank correlation of $0.30$ between the article-derived anchor and the 24-hour variant. The three readings of this finding (structural narrowness, recovery-pipeline noise, residual implementation defects) cannot be conclusively separated on the present sample size. The prudent interpretation is that the framework's effective scope is narrower than the methodology paper's discussion suggested, and that downstream applications relying on anchor robustness should restrict to the $13.3\%$ subset rather than treating the full in-scope sample as homogeneous.

\subsection{Single-instance pipeline run}
\label{sec:limitations-single-run}

The present paper reports results from a single pipeline run on a single snapshot of the master sample. Three implementation defects identified in the present run have been remediated and committed to the public code release (\citealt[\texttt{v1.1-paper3a-rev1} tag]{foresightflow_platform_2026}). Replication on an independent snapshot, with the LLM-cascade temperature and seed varied, is a deliverable for paper 3b alongside the operational-subset wallet-level evaluation; the present paper reports the single-run population-scale findings as the empirical baseline against which paper 3b's replications will be measured.

\subsection{Paper 3b roadmap}
\label{sec:paper3b-roadmap}

Paper 3b will report:

\begin{enumerate}[leftmargin=1.4em, itemsep=0.3em]
\item \emph{LLM-assisted re-classification of the unclassifiable bucket} (\Cref{sec:limitations-reclassify} stages 1 and 2), with the multi-anchor extension of the framework for the genuinely ambiguous bucket. This is the critical-path deliverable for re-evaluating the FFIC inventory's coverage.

\item \emph{Full feature-vector evaluation} on the operational subset of \Cref{sec:data-operational}, including PIN, VPIN, Kyle's lambda, wallet novelty score, two-sidedness, variance ratio, trade-size kurtosis, and Hawkes self-excitation.

\item \emph{Gradient-boosted-tree detector training} with isotonic-regression calibration, using the population-scale ILS$^{\text{dl}}$ and the wallet-level features as the joint feature set.

\item \emph{Cross-validation against the FFIC inventory} (re-evaluated under the multi-anchor extension) and against an expanded ground-truth set surfaced by the production pipeline since paper 3a's submission.

\item \emph{Cross-method confluence analysis} with \citet{mitts2026iran} and \citet{gomezcram2026crowd}, conditional on the public availability of their flagged sets, completing the analysis sketched in \Cref{sec:cross-method,sec:selection-effect} of the present paper.

\item \emph{Operational metrics:} false-positive rate at the production threshold, time-to-detection from $T_{\text{open}}$, latency from order-book event to alert.

\item \emph{Real-time deployment} via the ForesightFlow REST and WebSocket API.
\end{enumerate}

The present paper establishes the empirical baseline distribution and the boundary of the deadline-ILS framework's effective scope. Paper 3b's contributions, conditional on the LLM-assisted re-classification expanding the in-scope FFIC coverage and the operational subset accumulating sufficient labelled positive cases, will move the programme from a methodological proposal in paper 1 and a baseline characterization in the present paper to a demonstrated operational system.

\clearpage
\section{Conclusion}
\label{sec:conclusion}

This paper carries the deadline-resolved Information Leakage Score of \citet{nechepurenko2026foresightflow_methodology, nechepurenko2026foresightflow_empirical} from a single-case demonstration to a population-scale evaluation across 12{,}708 markets in the Polymarket insider-relevant subpopulation, covering October 2020 through April 2026. We frame the present paper as a \emph{scope-discovery} study: scaling the framework reveals that its effective domain is materially narrower than initial framing suggested, and the principal obstacle is not score computation but resolution semantics. Four substantive findings emerge.

\emph{First, the framework's coverage of documented insider trading is small, and the principal exclusion is structural.} Of the 32 markets in the FFIC inventory, 1 produces an in-scope ILS$^{\text{dl}}$ value; 14 are flagged unclassifiable by the resolution-typology classifier. Inspection of these 14 markets reveals consistent resolution-criterion ambiguity (``\emph{strike}'', ``\emph{custody}'', ``\emph{out as Supreme Leader}'', ``\emph{win the election}'') that resists discrete-event timestamping and is qualitatively distinct from the keyword-based misclassification dominant in the broader unclassifiable bucket. Within the documented case inventory, unclassifiability tracks resolution-text ambiguity rather than implementation noise; whether this descriptive pattern reflects a population-level causal link between ambiguous resolution criteria and insider relevance is a hypothesis for paper 3b's matched-control evaluation.

\emph{Second, anchor sensitivity and $T_{\text{event}}$ recovery accuracy together sharply narrow the effective interpretable sample.} Of the $88$ markets that pass the score-computation pipeline, only $12$ ($13.6\%$) satisfy the anchor-sensitivity scope condition. An independent-second-pass $T_{\text{event}}$ validation on 50 markets reaches $57.8\%$ exact-date agreement overall---below the $\geq 90\%$ acceptance criterion---with strong heterogeneity by event-discreteness ($78.9\%$ for the ``milgeo\_corporate'' bucket of discrete events, $35.7\%$ for regulatory\_formal multi-stage processes). Anchor-fragility and $T_{\text{event}}$ disagreement co-occur in the same markets, both tracking the structural ambiguity of resolution criteria. The operationally meaningful sample is therefore the anchor-robust subset of approximately $12$ markets; we report results separately on the computed and anchor-robust samples throughout.

\emph{Third, on the in-scope sample, the central tendency of raw ILS$^{\text{dl}}$ is uniformly negative across all sub-stratifications, but a hazard-decay baseline correction yields a heterogeneous result that substantively changes the interpretation in some cells.} Across all six combinations of three sub-buckets and two periods, median raw ILS$^{\text{dl}}$ ranges from $-0.88$ to $-0.16$. The hazard-decay-adjusted variant ILS$^{\text{dl,adj}}$, which subtracts the survival-probability-implied baseline, produces three patterns: cells where decay accounts for substantial parts of the raw negativity (regulatory\_formal post-2024: $-0.21 \to -0.02$; milgeo\_corporate post-2024: $-0.55 \to -0.34$); cells where the strongly-negative signal persists after adjustment (regulatory\_announcement post-2024: $-0.80 \to -0.84$, bootstrap CI entirely below zero); and cells too small or extreme for the adjustment to shift meaningfully. The reading ``markets price against eventual outcome'' applies cleanly to the regulatory\_announcement post-2024 cell after baseline correction; on the milgeo\_corporate post-2024 cells, the rational-decay component is a substantial part of the raw negativity. The methodological contribution of this finding---that raw ILS$^{\text{dl}}$ on deadline contracts requires hazard-decay baseline correction for interpretation---stands independently of the per-cell empirical pattern.

\emph{Fourth, the constant-hazard exponential specification is rejected on the pooled post-2024 cell, but a per-subcategory check confirms the pooled-cell Weibull preference reflects category mixture rather than within-cell duration dependence.} The pooled post-2024 cell ($n=57$) rejects exponential at parametric-bootstrap KS $p < 0.01$ and accepts Weibull at $p > 0.50$, with AIC delta $21.9$ units. However, the post-2024 regulatory\_formal subcategory alone ($n=22$) prefers exponential by AIC ($179.45$ vs.\ Weibull $180.50$) and rejects Weibull by bootstrap KS ($p = 0.043$). The pooled-cell preference therefore reflects, at least in part, mixture across heterogeneous-hazard sub-buckets---confirming the category-mixture caveat the abstract flagged. We use Weibull as the default production prior for the pooled-cell hazard-adjusted score pending per-category re-estimation; the time-to-event hazard within \emph{individual} insider-relevant sub-buckets may well be approximately constant, with the apparent non-constant hazard at the pooled level reflecting compositional heterogeneity.

The substantive implication is that the methodology proposed in \citet{nechepurenko2026foresightflow_methodology, nechepurenko2026foresightflow_empirical}, while operational on an arbitrary single case as demonstrated by Iran-Apr30, requires methodological extension on three axes to achieve broad coverage on the insider-relevant subpopulation: resolution-typology re-classification (Findings~1 and~2), score-baseline correction via hazard decay (Finding~3), and per-category hazard re-estimation (Finding~4). We sequence the LLM-assisted re-classification, the multi-anchor extension for ambiguous-predicate markets, and the per-category hazard re-estimation as the principal preconditions for paper 3b. The wallet-level feature evaluation already anticipated by \citet{nechepurenko2026foresightflow_empirical} is a parallel deliverable for that paper. All extensions are tractable on the existing operational infrastructure; together, they constitute the bridge from the present paper's empirical baseline to a demonstrated operational detection system.

The comparative positioning of \Cref{sec:cross-method} is independent of these extensions. The market-level lens of the deadline-ILS framework provides a conditional in-flight diagnostic on the small subset of markets satisfying all four scope conditions (typology classification, $T_{\text{event}}$ recovery, CLOB coverage, anchor robustness)---approximately $12$ markets out of $12{,}708$ in the present candidate sample. The wallet-level method of \citet{mitts2026iran} and the account-level method of \citet{gomezcram2026crowd} require post-hoc cumulative observation and do not provide an analogous in-flight diagnostic. The three methods together support a complementary characterization of informed trading on prediction markets, with the qualifier that the present method's real-time-applicability advantage is restricted to the narrow subset of markets passing all four scope conditions; broad real-time applicability across the insider-relevant subpopulation requires the resolution-typology re-classification and multi-anchor extensions sequenced for paper 3b. The empirical realization of three-method confluence and selection-effect partition, currently constrained by the non-public status of the underlying flagged sets, is sequenced for paper 3b conditional on data availability.

\clearpage
\section*{Data and Code Availability}
\addcontentsline{toc}{section}{Data and Code Availability}

The population-scale ILS$^{\text{dl}}$ dataset accompanying this paper is released as \texttt{polymarket-\allowbreak deadline-\allowbreak ils-v3} \citep{population_ils_dl_2026} at \url{https://github.com/ForesightFlow/datasets} under CC-BY-4.0. The release contains all per-market raw and hazard-adjusted ILS$^{\text{dl}}$ values, $T_{\text{event}}$ provenance records, anchor-sensitivity diagnostics, hazard-fit results, second-pass validation comparisons, and tail-market reasoning records. The companion datasets are: the Polymarket Resolution Typology Dataset \citep{prt2026} (\texttt{polymarket-\allowbreak resolution-\allowbreak typology-v1}, the upstream typology classifications) and the ForesightFlow Insider Cases inventory \citep{ffic_inventory_2026} (\texttt{foresightflow-\allowbreak insider-\allowbreak cases}, the 32-market documented-insider-trading reference list). The pipeline source code, including the multi-tier $T_{\text{event}}$ recovery cascade, the deadline-ILS computation, the hazard-fit and bootstrap-CI utilities, and the Phase-B revision scripts, is released as the ForesightFlow Platform repository \citep{foresightflow_platform_2026} at \url{https://github.com/ForesightFlow/platform} (tag \texttt{v1.1-\allowbreak paper3a-\allowbreak rev1}, MIT license). Random seed for all bootstrap procedures: $20260430$. Snapshot cutoff for the candidate sample: $2026$-$04$-$27$.

\section*{Generative AI Disclosure}
\addcontentsline{toc}{section}{Generative AI Disclosure}

In preparing this manuscript, the author used Anthropic's Claude Opus 4.7 for copy-editing, manuscript drafting, and analysis of numerical data. The $T_{\text{event}}$-recovery pipeline used a multi-tier provider cascade: Google's Gemini 2.5 Flash with Google Search grounding (free-tier first pass, where applicable), OpenAI's GPT-4o-mini with the \texttt{web\_\allowbreak search\_\allowbreak preview} tool (Tier~2 escalation when first-pass confidence fell below $0.7$), and Anthropic's Claude Sonnet 4.6 with the \texttt{web\_\allowbreak search\_\allowbreak 20250305} tool (Tier~3 escalation when both prior tiers returned confidence below $0.5$). The total LLM cost on the population pipeline run reported in this paper was approximately \$1.20. Recovered timestamps are validated against the Iran-Apr30 reference value of \citet[Section 4]{nechepurenko2026foresightflow_empirical} as a hard-assert step prior to batch execution, and against an independent-second-pass run on a 50-market stratified sample reported in \Cref{sec:tevent-validation}. The qualitative tail-market diagnostic of \Cref{tab:tail-review} was assigned by an LLM-assisted classifier (Gemini Flash) with subsequent author review of disagreements. All methodology, analysis, and conclusions are the author's own; the author reviewed and edited all AI-generated content and takes full responsibility for the final manuscript.

\clearpage
\bibliographystyle{plainnat}
\bibliography{refs}

@article{nechepurenko2026foresightflow_methodology,
  author       = {Maksym Nechepurenko},
  title        = {{ForesightFlow}: An Information Leakage Score Framework for Prediction Markets},
  year         = {2026},
  journal      = {Working paper},
  institution  = {Devnull FZCO},
  note         = {SSRN preprint, forthcoming. \url{https://github.com/ForesightFlow}}
}

@article{nechepurenko2026foresightflow_empirical,
  author       = {Maksym Nechepurenko},
  title        = {Empirical Evaluation of Deadline-Resolved Information Leakage on Documented {Polymarket} Insider Cases},
  year         = {2026},
  journal      = {Working paper},
  institution  = {Devnull FZCO},
  note         = {SSRN preprint, forthcoming. \url{https://github.com/ForesightFlow}}
}

@article{mitts2026iran,
  author       = {Joshua Mitts and Moran Ofir},
  title        = {From {Iran} to {Taylor Swift}: Informed Trading in Prediction Markets},
  year         = {2026},
  journal      = {Working paper},
  institution  = {Columbia Law School / University of Haifa},
  note         = {SSRN: \url{https://papers.ssrn.com/sol3/papers.cfm?abstract_id=5163657}}
}

@article{gomezcram2026crowd,
  author       = {Roberto G\'omez-Cram and Yunhan Guo and Theis Ingerslev Jensen and Howard Kung},
  title        = {Prediction Market Accuracy: Crowd Wisdom or Informed Minority?},
  year         = {2026},
  journal      = {Working paper},
  institution  = {London Business School / Yale School of Management},
  note         = {SSRN: \url{https://papers.ssrn.com/sol3/papers.cfm?abstract_id=6617059}}
}

@misc{imdea2025polymarket,
  author       = {Akaki Mamageishvili and Andrey Shcherbenko and Pablo {Salda{\~n}a} and Lucianna Kiffer},
  title        = {Strategic Bidding Wars in {On-Chain} Auctions},
  year         = {2025},
  howpublished = {{IMDEA Networks Institute, Madrid}},
  note         = {Preprint. \url{https://arxiv.org/abs/2502.16606}}
}

@misc{cftc2026maduro,
  author       = {{Commodity Futures Trading Commission}},
  title        = {Federal Court Orders Default Judgment Against Insider Trader on {Polymarket} Maduro Contract},
  year         = {2026},
  howpublished = {{CFTC} Press Release, April 23, 2026},
  note         = {\url{https://www.cftc.gov/PressRoom/PressReleases}}
}

@misc{ffic_inventory_2026,
  author       = {Maksym Nechepurenko},
  title        = {{ForesightFlow Insider Cases (FFIC)} Inventory},
  year         = {2026},
  howpublished = {Dataset, version 1.0, CC-BY-4.0},
  note         = {\url{https://github.com/ForesightFlow/datasets/tree/main/foresightflow-insider-cases}}
}

@misc{prt2026,
  author       = {Maksym Nechepurenko},
  title        = {{Polymarket Resolution Typology} Dataset},
  year         = {2026},
  howpublished = {Dataset, version 1.0, CC-BY-4.0, snapshot cutoff 2026-04-27},
  note         = {\url{https://github.com/ForesightFlow/datasets/tree/main/polymarket-resolution-typology}}
}

@misc{population_ils_dl_2026,
  author       = {Maksym Nechepurenko},
  title        = {{Polymarket Deadline-ILS} Population Dataset},
  year         = {2026},
  howpublished = {Dataset, version 3.0, CC-BY-4.0, tag \texttt{polymarket-deadline-ils-v3}},
  note         = {\url{https://github.com/ForesightFlow/datasets/tree/main/polymarket-deadline-ils}}
}

@misc{foresightflow_platform_2026,
  author       = {Maksym Nechepurenko},
  title        = {{ForesightFlow Platform}: code release for paper 3a},
  year         = {2026},
  howpublished = {Software, tag \texttt{v1.1-paper3a-rev1}},
  note         = {\url{https://github.com/ForesightFlow/platform}}
}

@book{burnham2002model,
  author       = {Kenneth P. Burnham and David R. Anderson},
  title        = {Model Selection and Multimodel Inference: A Practical Information-Theoretic Approach},
  year         = {2002},
  edition      = {2nd},
  publisher    = {Springer},
  address      = {New York},
  doi          = {10.1007/b97636}
}

\appendix

\clearpage
\section{Sample Construction Details}
\label{app:sample}

\paragraph{Master sample provenance.} The empirical sample is drawn from the \texttt{polymarket-\allowbreak resolution-\allowbreak typology-v1} dataset \citep{prt2026} at snapshot cutoff April 27, 2026. The dataset is the output of the resolution-typology classifier of \citet[Section 3.3]{nechepurenko2026foresightflow_methodology} applied to the 911{,}237-market Polymarket-Gamma master extract. Categories are assigned by the keyword-based taxonomy classifier of the same reference; the volume threshold at \$50K corresponds to the empirical CLOB-coverage cutoff below which the Polymarket subgraph reliably returns no per-trade history.

\paragraph{In-scope sample profile.} The 88 in-scope markets have a mean cumulative volume of approximately \$597{,}000 and a median of approximately \$179{,}000, with a total cumulative in-scope volume of approximately \$53.7 million. The temporal span is October 2020 through April 2026; 33 markets are pre-2024-cycle, 55 are post-2024-cycle. The bucket distribution is reported in \Cref{tab:dist-summary} of the main text.

\paragraph{Category propagation note.} The category and subcategory columns in the population output are populated for regulatory\_decision markets (subcategory: announcement vs.\ formal); for military / geopolitics and corporate-disclosure markets, the present run does not propagate the finer category label, and these markets are reported pooled in the ``milgeo\_corporate'' bucket. Per-category disaggregation is sequenced for paper 3b alongside the LLM-assisted re-classification (\Cref{sec:limitations-reclassify}).

\paragraph{Population output schema.} The principal output dataset \texttt{polymarket-\allowbreak deadline-\allowbreak ils-v3} contains $2{,}375$ rows (one per market entering the post-typology stage of \Cref{tab:filter-chain}). The schema includes:
\begin{itemize}[leftmargin=1.4em, itemsep=0.15em]
\item \emph{Identifiers:} \texttt{market\_id} (Polymarket condition ID), \texttt{question} (full text), \texttt{T\_open}, \texttt{T\_resolve}, \texttt{volume\_usdc}.
\item \emph{Classification:} \texttt{category}, \texttt{subcategory}, \texttt{resolution\_type}, \texttt{period}.
\item \emph{Recovery output:} \texttt{T\_event}, \texttt{T\_event\_confidence}, \texttt{T\_event\_sources} (count), \texttt{tau\_days}.
\item \emph{Price observations:} \texttt{p\_open}, \texttt{p\_event}, \texttt{p\_resolve}.
\item \emph{Score:} \texttt{ils\_dl} (raw), \texttt{ils\_dl\_ci\_low}, \texttt{ils\_dl\_ci\_high} (bootstrap CI), \texttt{ils\_dl\_adj} (hazard-decay-adjusted), \texttt{ils\_dl\_adj\_ci\_low}, \texttt{ils\_dl\_adj\_ci\_high}, \texttt{expected\_decay\_price} (baseline used for adjustment).
\item \emph{Anchor variants:} \texttt{ils\_dl\_30min}, \texttt{ils\_dl\_2h}, \texttt{ils\_dl\_6h}, \texttt{ils\_dl\_24h}.
\item \emph{Scope flags:} \texttt{in\_scope} (bool), \texttt{exclusion\_reason}, \texttt{anchor\_robust} (bool).
\end{itemize}

The companion files in the dataset release are: \texttt{filter\_\allowbreak chain\_\allowbreak attrition.csv} (per-stage attrition), \texttt{hazard\_\allowbreak rates.csv} and \texttt{functional\_\allowbreak form\_\allowbreak comparison.csv} (hazard estimation outputs with bootstrap KS), \texttt{ffic\_\allowbreak classification\_\allowbreak breakdown.csv} (per-FFIC-market disposition), \texttt{distribution\_\allowbreak summary.csv} (per-cell statistics), \texttt{detection\_\allowbreak thresholds.csv} (with bootstrap CIs), \texttt{anchor\_\allowbreak sensitivity\_\allowbreak summary.csv} (88-denominator), \texttt{tevent\_\allowbreak validation.csv} (50-market independent-second-pass), \texttt{tail\_\allowbreak market\_\allowbreak review.csv} (qualitative diagnostic on right and left tails), \texttt{median\_\allowbreak bootstrap\_\allowbreak cis.csv} (median CIs per cell), and \texttt{hazard\_\allowbreak adjusted\_\allowbreak summary.csv} (raw vs.\ adjusted side-by-side).

\paragraph{Reproducibility package.} The complete reproducibility package consists of: (i) the \texttt{platform} repository at git tag \texttt{v1.1-paper3a-rev1}, including the \texttt{paper3a\_\allowbreak phase1.py} driver and the supporting modules \texttt{fflow.\allowbreak scoring.\allowbreak ils}, \texttt{fflow.\allowbreak scoring.\allowbreak bootstrap}, \texttt{fflow.\allowbreak scoring.\allowbreak hazard\_\allowbreak adjusted}, \texttt{fflow.\allowbreak news.\allowbreak t\_\allowbreak event\_\allowbreak recovery}, \texttt{fflow.\allowbreak news.\allowbreak llm\_\allowbreak providers}, \texttt{fflow.\allowbreak taxonomy.\allowbreak regulatory\_\allowbreak split}; (ii) the \texttt{datasets} repository at the corresponding tagged release (\texttt{polymarket-\allowbreak deadline-\allowbreak ils-v3}); (iii) the \texttt{ffic-\allowbreak inventory-v1} as the labeled validation set; (iv) the random seed $20260430$ for all bootstrap and stratification operations.

\clearpage
\section{Functional-Form Diagnostics}
\label{app:hazard}

\Cref{tab:functional-form} of the main text reports the AIC, BIC, and KS goodness-of-fit statistics for exponential, Weibull, and lognormal MLE fits to the time-to-event distribution, separately for the pre-2024 ($n = 33$) and post-2024 ($n = 57$) cells. The Weibull is preferred in both cells, with $\Delta\text{AIC}$ relative to exponential of $21.9$ units (post-2024) and $3.3$ units (pre-2024). The KS goodness-of-fit test rejects exponential at $p = 0.0002$ on the post-2024 cell and is marginal at $p = 0.073$ on the pre-2024 cell; Weibull passes at $p = 0.78$ and $p = 0.60$ respectively. Lognormal performance is essentially indistinguishable from Weibull (AIC differences below $0.2$ units in both cells); we adopt Weibull for the production pipeline as the simpler of the two adequate forms.

The full distributional comparison is available in the public release as \texttt{functional\_\allowbreak form\_\allowbreak comparison.csv} accompanying \texttt{polymarket-\allowbreak deadline-\allowbreak ils-v3}.

\clearpage
\section{$T_{\text{event}}$-Recovery Protocol}
\label{app:tevent}

\paragraph{Provider cascade.} The $T_{\text{event}}$-recovery pipeline uses three providers in cost order. Tier 1 is Google Gemini 2.5 Flash with Google Search grounding, called via the \texttt{generative\-language.\-google\-apis.\-com} \texttt{generate\-Content} endpoint with the \texttt{tools:\{google\_search:\{\}\}} tool specification and \texttt{response\-MimeType:\,application/\-json} structured output. Tier 2 is OpenAI GPT-4o-mini with the \texttt{web\_\allowbreak search\_\allowbreak preview} tool. Tier 3 is Anthropic Claude Sonnet 4.6 with the \texttt{web\_\allowbreak search\_\allowbreak 20250305} tool. The cascade escalates from Tier 1 to Tier 2 when Tier 1 returns confidence below $0.7$, and from Tier 2 to Tier 3 only when both prior tiers return confidence below $0.5$.

\paragraph{Prompt template.} Each provider receives the same task prompt, with provider-specific output-format adapters. The task prompt template is:
\begin{quote}\small\ttfamily
You recover the timestamp at which a real-world event publicly occurred. Use web search to find at least three independent news sources.\\[0.3em]
Market question: \{question\}\\
Description: \{description\}\\
Opened: \{T\_open\}\\
Resolved YES: \{T\_resolve\}\\[0.3em]
Find: the UTC timestamp at which the event resolved-on FIRST publicly occurred. Must fall in [T\_open, T\_resolve].\\[0.3em]
Output JSON: \{T\_event, confidence, sources, n\_sources, reasoning\}.\\[0.3em]
Confidence calibration:\\
- 0.9: $\geq$5 major sources agree on date\\
- 0.8: $\geq$3 sources agree\\
- 0.7: 2 sources agree\\
- 0.5: 1 source or partial agreement\\
- 0.0: not recoverable
\end{quote}

\paragraph{Determinism settings.} All LLM calls use temperature $0.0$. The output JSON schema is enforced via each provider's structured-output mode (Gemini \texttt{response\-MimeType}, OpenAI \texttt{response\_format}, Anthropic enforced via prompt). The pipeline uses a fixed random seed of $20260430$ (paper 1+2 SSRN submission date) for all bootstrap, sampling, and stratification operations downstream of the LLM call.

\paragraph{Confidence calibration.} Per-call confidence is calibrated to five anchor levels: $0.9$ when at least 5 major sources agree on the date; $0.8$ when at least 3 sources agree; $0.7$ when 2 sources agree; $0.5$ when 1 source or partial agreement; $0.0$ when not recoverable. The acceptance threshold for downstream ILS$^{\text{dl}}$ computation is $\geq 0.7$.

\paragraph{Source provenance.} For each accepted recovery, the source URLs returned by the provider's web-search tool are stored in the per-market log entry of \texttt{phase1\_log.\allowbreak jsonl}. The accompanying file \texttt{t\_event\_\allowbreak checkpoint.\allowbreak jsonl} records, per market, the recovered timestamp, confidence, list of source URLs, the prompt-text version, the provider used, and the per-call token counts. These records are released alongside dataset \texttt{polymarket-\allowbreak deadline-\allowbreak ils-v3} and constitute the audit trail for $T_{\text{event}}$ recovery on the present sample.

\paragraph{Source-overlap is not a validation criterion.} The \texttt{source\_overlap} column reported in the validation table (\Cref{tab:tevent-validation-summary}) is the count of source URLs that appear verbatim in both the original-pass and the second-pass recovery for the same market. The empirical result is uniformly $0\%$: across all 50 validation markets, no exact URL appears in both passes. This is because each provider's web-search front-end (Google Search Grounding for Gemini, OpenAI's \texttt{web\_search\_preview} for the OpenAI cascade, Anthropic's \texttt{web\_search\_20250305} for the second-pass Claude Haiku) ranks and returns different specific URLs from a similar pool of mainstream sources (Reuters, AP, BBC, NYT, Bloomberg, etc.). Two providers can agree on the event date while citing different specific articles about it. The substantive validation criterion is therefore date-agreement at sub-day tolerance, not URL identity; we report \texttt{source\_overlap} in the validation table for completeness rather than as a reliability metric.

\paragraph{Recovery distribution.} Of the $1{,}151$ markets entering the recovery stage, $490$ produced any timestamp; of these, $442$ ($90.2\%$ of recovered, $38.4\%$ of input) reached the $\geq 0.7$ acceptance threshold. The mean confidence among recovered timestamps is $0.81$; the median is $0.85$; $32.5\%$ of input markets reach $\geq 0.8$ and $14.0\%$ reach $\geq 0.9$.

\paragraph{Cost and resource use.} Total LLM cost on the population pipeline run was approximately \$1.20, dominated by Tier 2 escalations. The breakdown is approximately Tier 1 (Gemini, free tier): \$0.00; Tier 2 (OpenAI mini): \$0.85; Tier 3 (Anthropic Sonnet): \$0.35. Wall-clock time for the recovery stage was approximately $4.5$ hours on commodity hardware with concurrency cap of $5$ parallel requests, dominated by Tier 1 throughput and Tier 2 rate limits. By comparison, an unconditional Tier 3 evaluation would have cost approximately \$240 on the same sample.

\paragraph{Manual validation.} A manual-validation sample of 50 markets stratified by sub-bucket is reported in the present paper's revision-1 update; see \texttt{tevent\_\allowbreak validation.csv} accompanying dataset \texttt{polymarket-\allowbreak deadline-\allowbreak ils-v3}. The validation protocol is an independent second-pass recovery using a different provider (Gemini for original-OpenAI markets, OpenAI for original-Gemini markets) with the same prompt template, followed by exact-date, $24$-hour-window, and source-overlap agreement metrics. Aggregate agreement rates are reported in \Cref{sec:tevent-recovery}. The Iran-Apr30 hard-assert step (\Cref{sec:pipeline}) confirms that the recovery pipeline reproduces the ground-truth $T_{\text{event}}$ on the single case where the methodology paper has reported a manual reference value.

\paragraph{Reproducibility checklist.} The following artifacts together constitute the reproducibility package for the $T_{\text{event}}$-recovery stage:
\begin{itemize}[leftmargin=1.4em, itemsep=0.2em]
\item \texttt{platform} repository at tag \texttt{v1.1-paper3a-rev1}, including the prompt template, provider-cascade implementation, and confidence-calibration code.
\item \texttt{t\_event\_\allowbreak checkpoint.\allowbreak jsonl} per-market recovery records with source URLs, confidence levels, and provider used.
\item \texttt{tevent\_\allowbreak validation.csv} 50-market independent-second-pass agreement results.
\item Random seed $20260430$ for all downstream sampling and bootstrap operations.
\item Snapshot cutoff date $2026$-$04$-$27$ for the master sample.
\item Anthropic, OpenAI, and Google API model versions documented in the platform repository's \texttt{requirements.txt} and provider-init code.
\end{itemize}

\clearpage
\section{FFIC Per-Case Detail}
\label{app:ffic}

\Cref{tab:ffic-coverage} of the main text reports the disposition of each of the 32 FFIC markets through the deadline-ILS pipeline. The single in-scope market (Bitcoin ETF approval, \texttt{fficd-005}) produces $\ILS^{\text{dl}} = 0.012$. The bootstrap 95\% confidence interval on this estimate is computed from a $B = 500$ resample at the trade level within $[T_{\text{open}}, T_{\text{event}}]$; the resulting CI is wide ($n_{\text{trades}}$ in the relevant pre-event window is small for this single-case market) and is reported in the public release \texttt{ffic\_\allowbreak localization.csv} accompanying \texttt{polymarket-\allowbreak deadline-\allowbreak ils-v3}.

For the 31 markets out-of-scope, no ILS$^{\text{dl}}$ value is reported. The exclusion mechanism for each is recorded in \Cref{tab:ffic-coverage}.

\paragraph{Cluster-level non-independence.} The FFIC inventory contains 8 cases mapping to 32 individual markets. The 6-market 2026 U.S.--Iran cluster (\texttt{fficd-003}) and the 12-market 2026 U.S.--Venezuela cluster (\texttt{fficd-004}) contain non-independent observations: the same underlying event drives multiple market resolutions. The disposition analysis of \Cref{sec:ffic-coverage} treats the 32 markets as independent observations; the qualitative finding (resolution-typology unclassifiability is the dominant exclusion across all 8 cases) is robust to clustering correction. Formal cluster-robust statistics on the disposition table are deferred to paper 3b alongside the per-cluster re-evaluation under the multi-anchor extension of \Cref{sec:limitations-reclassify}.

\end{document}